\documentclass[11pt,fullpage]{article}

\usepackage{ulem}

\usepackage{latexsym}
\usepackage{multirow}
\usepackage{slashbox}
\usepackage{color}
\usepackage[toc]{appendix}
\usepackage[pdftex]{graphicx}

\usepackage{amsmath}
\usepackage{amssymb}
\usepackage{amsthm}
\usepackage{amsfonts}
\usepackage{url}

\newcommand {\bb}[1]{{\mathbb #1}}

\renewcommand{\baselinestretch} {1}

	\makeatletter 
\def\singlespace{\def\baselinestretch{1}\@normalsize}

\newtheorem{theorem}{Theorem}
\newtheorem{lemma}{Lemma}

\newtheorem{corollary}{Corollary}
\newtheorem{proposition}{Proposition}
\newtheorem{remark}{Remark}

\newcommand{\be}{\begin{equation}}
\newcommand{\ee}{\end{equation}}
\newcommand{\beqn}{\begin{eqnarray}}
\newcommand{\eeqn}{\end{eqnarray}}
\newcommand{\beqns}{\begin{eqnarray*}}
\newcommand{\eeqns}{\end{eqnarray*}}

\newcommand{\fr}[1]{(\ref{#1})} % formula
\newcommand{\lkr}{\left(}  %left kruglaya
\newcommand{\rkr}{\right)} %right kruglaya
\newcommand{\lkv}{\left[}  %left kvadratnaya
\newcommand{\rkv}{\right]} %right kvadratnaya
\newcommand{\lfi}{\left\{} %left figurnaya
\newcommand{\EE}{\ensuremath{{\mathbb E}}}

\newcommand{\ZZ}{\ensuremath{{\mathbb Z}}}

\newcommand{\wind}{\mbox{wind}}
\newcommand{\Tr}{\mbox{Tr}}

\newcommand{\pen}{\mbox{pen}}
\newcommand{\Var}{\mbox{Var}}
\newcommand{\AIF}{\mbox{AIF}}

\newcommand{\om}{\omega}

\newcommand{\bA}{\mbox{$\mathbf A$}}

\newcommand{\bId}{\mbox{\boldmath $Id$}}

\newcommand{\bG}{\mbox{$\mathbf G$}}
\newcommand{\bQ}{\mbox{$\mathbf Q$}}
\newcommand{\bs}{\mbox{$\mathbf s$}}
\newcommand{\bt}{\mbox{$\mathbf t$}}
\newcommand{\bq}{\mbox{$\mathbf q$}}

\newcommand{\bof}{\mbox{$\mathbf f$}}
\newcommand{\bJ}{\mbox{$\mathbf J$}}

\newcommand{\bzeta}{\mbox{\mathversion{bold}$\zeta$}}

\newcommand{\bOm}{\mbox{\mathversion{bold}$\Omega$}}

\newcommand {\hfapr}{\widehat f_{\mathrm{APR}}}

\newcommand{\whm}{\widehat{m}}

\newcommand{\mathM}{{\mathcal M}}

\graphicspath{%
    {converted_graphics/}% inserted by PCTeX
    {/}% inserted by PCTeX
}
\begin{document}

\title{\Large{\bf Laplace deconvolution on the basis of   time domain data  and its application to Dynamic Contrast Enhanced imaging}}

\author{
\large{{\sc Fabienne Comte$^{1,2}$}, {\sc Charles-A. Cuenod$^{1,3}$ },   {\sc Marianna Pensky$^4$}}\\
\large{and {\sc Yves Rozenholc$^{1,2,5}$}}
  \\ \\
   Universit\'{e}  Paris Descartes$^1$, MAP5, UMR CNRS 8145$^2$, LRI INSERM U970 PARCC-HEGP$^3$,\\    
University of Central Florida$^4$ and INRIA Saclay Ile-de-France  {\sc select}$^5$}

\date{}

\maketitle

\noindent \noindent
Corresponding author: F. Comte, 
Universit\'{e}  Paris  Descartes, MAP5, UMR CNRS 8145, France\\
{\em fabienne.comte@parisdescartes.fr}\\

% \vspace{1cm}
\begin{abstract}
In the present paper we consider the problem of Laplace deconvolution
with noisy discrete non-equally spaced observations on a finite time interval. 
We propose a new method for Laplace deconvolution which is based on
expansions of the convolution kernel, the unknown function and the observed
signal over  Laguerre functions basis (which acts as a surrogate eigenfunction 
basis  of the Laplace convolution operator) using regression setting.
The expansion results in a small system of linear equations with the matrix of the
system being triangular and Toeplitz. Due to this triangular structure, there is a common 
number $m$ of  terms in the function expansions to control, which is realized  via complexity penalty.
The advantage of this methodology is that it leads to very fast computations, 
produces no boundary effects due to extension at zero and cut-off at $T$ and
provides an estimator  with the risk within a logarithmic factor
of $m$ of the oracle risk. We emphasize that, in the present paper,  
we consider the true observational model with possibly non-equispaced observations
which are available on a finite interval of length $T$ which appears 
in  many  different contexts, and account for the bias associated
with this model  (which is not present in the case of $T\rightarrow\infty$).

 The study is motivated by perfusion imaging using a short injection of contrast agent, a procedure  which
 is   applied for medical assessment of   micro-circulation within tissues such as cancerous tumors.
Presence of a tuning parameter $a$ allows to choose the most advantageous time units, so
that both the kernel and the unknown right hand side of the equation are well represented for
the   deconvolution. 
 The methodology is illustrated by an extensive simulation study and a real data example
which confirms that the proposed technique is fast, efficient, accurate,
usable from a practical point of view and  very  competitive.
\end{abstract}

%%%%%%%%%%%%%%%%%%%%%%%%%%%%%%%%%%%%%%%%%%%%%%%%%%%%%%%%%%%%%%%%%%%%%%%%%%%%%%%%%%%

%\vspace{1cm}

\bibliographystyle{plain}

\pagestyle{plain}

\section{Introduction}
\label{sec:intro}
\setcounter{equation}{0}

 Consider the Laplace convolution model 
\be \label{eq:model}
y(t_i) = \int_0^{t_i} g(t_i - \tau) f(\tau) d\tau + \sigma \varepsilon_i, \quad i=1, \ldots, n,
\ee
 which is a discrete   noisy version of the linear Volterra equation of the first kind 
\be \label{eq:Volterra}
q(t) = \int_0^t g(t - \tau) f(\tau) d\tau = \int_0^t g(\tau) f(t-\tau) d\tau, \quad t \geq 0,
\ee
where function $g$ is considered to be known, $f$ is a function of interest,
measurements $y(t_i)$ are taken at points $0 \leq t_1 \leq ... \leq t_n \leq T < \infty$, and the errors
$\varepsilon_i$ are i.i.d.  sub-Gaussian random variables with
$\EE \varepsilon_i =0$ and $\Var (\varepsilon_i) = 1$ (see 
Vershynin (2012) for definition and discussion of sub-Gaussian random variables).

The study is motivated by   high frequency perfusion imaging 
 such as Dynamical Contrast Enhanced (DCE) imaging using either Computerized Tomography (DCE-CT),
% (where $n\approx 30$ and $T \approx 4$  minutes), 
Magnetic Resonance Imaging (DCE-MRI) 
% (where $n\approx 100$  and $T \approx 12$  minutes) 
or Ultra Sound (DCE-US).
 %(where $n$ can be few hundreds and $T$ few minutes). 
Those techniques have a great potential in cancer treatments but suffer from 
the lack of robust quantification as  pointed out by Cao  (2011).
 Model \fr{eq:model} is also used for describing  time-resolved measurements in fluorescence spectroscopy
(see, e.g. Ameloot and Hendrickx (1983), Ameloot {\it et al.} (1984), Gafni {\it et al.}  (1975), 
 O'Connor {\it et al.} (1979) and also the monograph of Lakowicz  (2006) and references therein).

 We   solve the problem \fr{eq:model} in a non-asymptotic setting where both $n$ and $T$ are not large, 
and time instances are not equally spaced, corresponding to the medical set-up which necessarily controls 
the patient's exposure to radiation together with the duration of the whole exam and is limited by the acquisition technique.  
Our objective is to design a technique which performs well under those conditions in a sense that it has 
minimal or nearly minimal possible error and  can be used when function $g$ is  only partially observed.
Therefore, we do not  replace equation \fr{eq:model} by an ``ideal''  white noise model which  immensely simplifies the issues  and
carry out error analysis in a practical observational set up. Below, we discuss existing methodologies and their limitations. 
\\
  
\subsection*{Solution by the Laplace transform}

The mathematical theory of (noiseless) convolution type Volterra equations
is well developed (see, e.g., Gripenberg  {\it et al.} 1990)
and the exact solution of  equation \fr{eq:Volterra} can be obtained through
Laplace transform.  However, direct application of Laplace transform for
discrete measurements faces serious conceptual and numerical problems.
The inverse Laplace transform is usually  found by application of tables of inverse Laplace
transforms, partial fraction decomposition or series expansion
(see, e.g., Polyanin and Manzhirov, 1998), neither of which is
applicable in the case of the discrete noisy version of Laplace deconvolution.

Numerical inversion of Laplace transform becomes unstable due to requirement 
of dividing it by the Laplace transform of function $g$.
Although the recently proposed   maximum entropy method of Mnatsakanov (2011) and Mnatsakanov and   
Sarkisian (2013)    works well for large sample sizes ($n \geq 500$), it dramatically deteriorates 
in the situations where $n$ is small  and one needs to recover solution of an ill-posed problem.

\subsection*{Fourier deconvolution}

Formally, by setting $g(t) = f(t) \equiv 0$ for $t<0$, equation
\fr{eq:Volterra} can be viewed as a particular case of the
Fourier convolution equation
\be \label{eq:Fredholm}
q(t) = \int_{-\infty}^\infty  g(t  - \tau) f(\tau) d\tau.
\ee
Discrete stochastic version of equation \fr{eq:Fredholm}
\be \label{eq:fourier_deconv}
y(t_i) = \int_{a}^b g(t_i - \tau) f(\tau) d\tau + \sigma \varepsilon_i, \;\;\;\; i=1,...,n,
\ee
known also as Fourier deconvolution problem,
has been extensively studied in the last thirty years
(see, for example, Carroll and Hall, 1988; Comte, Rozenholc and Taupin, 2006;
Delaigle, Hall and Meister, 2008;
Diggle and Hall, 1993; Fan, 1991; Fan and Koo, 2002; Johnstone {\it et al.},
2004; Pensky and Vidakovic, 1999;  Stefanski and Carroll, 1990, among others).

However, such an approach is  very misleading. To start with, although  in \fr{eq:Volterra} one has $0 \leq t \leq T$
with $T < \infty$, equation  $q(t) = \int_0^T g(t - \tau) f(\tau) d\tau$ {\bf is not} a Fourier convolution equation on the 
interval $[0,T]$, in the sense that application of the Fourier transform on an interval $[0,T]$ 
 {\bf does not} convert the integral  into  a  product  of the Fourier   transforms of $f$ and $g$ unless these      
functions $g$ and $f$  are  periodic on $[0,T]$, which is very unlikely to happen in applications.
Therefore, one has to apply the Fourier transform on the real line to equation \fr{eq:Fredholm}.
This application faces multiple obstacles: for small values of $n$ and $T$, inverse Fourier transform 
has poor precision since Fourier transform inherently operates on the whole real line 
and requires integration of highly oscillatory functions.
In addition, the true solution $f(\tau)$ may not vanish at $\tau =0$, which introduces an additional 
instability into the Fourier transform solution, due to a jump discontinuity of $f$ at zero.
Those difficulties, however, are not intrinsic to the problem and are entirely due to 
the usage of Fourier transform. Indeed, the concern of having measurements only for  $t \leq T$  
does not affect the Laplace deconvolution since it exhibits {\bf causality} property:
the values of $q(t)$ for $0\leq t \leq T$ depend on values of $f(t)$ for $0\leq t \leq T$
only and vice versa. Moreover, since function $f$ is considered only for $t\geq 0$, 
the issue of its discontinuity at zero does not arise.

\subsection*{Mathematical   approaches } 

  Several scientists  attempted to solve equation 
\fr{eq:model} using discretization  and then applying standard methodologies 
like the singular value decomposition (SVD) and the Tikhonov regularization (see, e.g.,  
Lamm (1996), Cinzori and Lamm (2000), and,  in the context of perfusion imaging,  
Ostergaard {\it et al.} (1996) and an extensive review in Fieselmann {\it et al.} (2011)).  
The shortcoming of these methods is that they are designed for a general linear inverse problem 
and do not take advantage of a particular form of the equation. In what follows, we compare our method with the SVD 
approach and  confirm that the latter one delivers very inferior estimators. In particular,
the  estimators exhibit  strong instabilities at $t=0$.

Methodology of  Ameloot and Hendrickx  (1983) is designed specifically for analysis of  fluorescence curves.
It relies on parametric presentation of the solution $f$ as a sum of exponential functions   
and requires the knowledge of the number of components. The approach is suitable only for the situation when 
the solution indeed has this parametric form and the number of components 
is small since the  exponential functions  are highly correlated.

The technique of Maleknejad {\it et al.} (2007) is based on the 
expansion of the solution over  the Haar wavelet basis.
The paper uses only the scaling parts of the Haar basis and finds
coefficients by  minimization of the discrepancy with the right-hand side.
The authors assume  the functions of interest to be piecewise constant and, 
hence, have an efficient representation in  Haar wavelet basis, 
which is not the case in our particular application.
Moreover, since the methodology is designed for exact measurements,
the authors  offer  no tools for model selection
and do  not provide statistical error bounds.  
Hence, despite being a general solution of the Laplace deconvolution problem, this method   
 is not a good option in the case of a small number of noisy irregularly spaced observations.

\subsection*{Statistical  approaches }

Unlike Fourier deconvolution, that has been intensively studied in statistical
literature (see references above), Laplace deconvolution
received very little  attention within statistical framework.
To the best of our knowledge, before 2010, only Dey, Martin and Ruymgaart (1998)
 tackled the statistical version of the problem. They  considered a noisy version of Laplace deconvolution
with a very specific kernel of the form $g(t)=be^{-at}$ and assumed that data are available
on the whole positive half-line (i.e.  $T= \infty$) and that smoothness of $f$ is known (i.e., the estimator is not adaptive).

Abramovich {\it et al.} (2013) studied the problem of Laplace deconvolution
based on discrete noisy data on a finite interval $[0, T]$. 
The idea of the method is to reduce the problem to
estimation  of the unknown regression function and its derivatives, 
using kernel method  with an adaptive choice of the bandwidth. 
The method has an advantage of reducing the Laplace deconvolution 
problem to a well studied nonparametric regression problem. Nevertheless, the shortcoming of the technique is that
it  is strongly dependent on {\bf the exact knowledge} of the kernel $g$  on the positive real line since it relies on the analytic 
inversion of the equation. In particular, it requires the knowledge of  the roots of   
the Laplace transform of the kernel $g$, leading to an extremely unstable estimator when 
exact analytic expression of the kernel is unknown
and $g$ is reconstructed using some measurements. 
Indeed, small change in  the observations of  $g$   produces significant  changes 
 in the roots and, hence, in the expression of the estimator. 
In addition, technique of Abramovich {\it et al.} (2013) requires meticulous  boundary correction.

\subsection*{Current methodology  }

The present paper offers a method which is designed to overcome limitations of
the previously developed techniques. The new methodology allows one
to use real-time data and is based on expansions of the kernel,
unknown function $f$ and the right-hand side in equation   \fr{eq:model} over the Laguerre functions basis.
As it was noticed before (see, e.g. Weeks (1966) or Lien {\it et al.} (2008)),
the Laguerre functions basis provides a surrogate eigenfunction   basis for the problem since
the expansions result  in a small system of linear equations with the matrix of the
system being  lower  triangular and Toeplitz.
The number of the terms in  the expansion  of the estimator is controlled via complexity penalty.

The technique does not require exact knowledge of the kernel since it is represented
by its Laguerre coefficients only, so, unlike  Abramovich {\it et al.} (2013),  it
can be easily applied in the case when the kernel $g$ is not known exactly  but is estimated from observations. 
The recent  Vareschi (2015)  paper, which  is built upon the first 
initial version of our manuscript (Comte {\it et al.} (2012)), makes this extension. 
However,  Vareschi (2015) considers a purely theoretical version of the model where one samples
Laguerre coefficients directly.  Contrary to this, in the present version of the paper, we provide
a true solution to the initial problem \fr{eq:model} and  estimate Laguerre coefficients 
 in the  regression set up. Note that, since Laguerre coefficients depend on the values of a function 
on $(0, \infty)$, estimation of the coefficients on the basis of limited data leads to an additional bias 
term which   can be made   smaller   than the squared bias and the variance of the 
penalized estimator. We provide an oracle inequality for the risk of the estimator and prove that, under 
mild assumptions on the model, the estimator is nearly optimal with the risk within $\log n$
factor of the minimal risk.

We would like to emphasize that, in the present paper, we examine the true observational model, where 
measurements are available only on a finite interval of length $T$ and are possibly non-equispaced.
This is a suitable description of data involved in, e.g,  high  frequency perfusion imaging as well as in 
other applications such as fluorescent spectroscopy.
To the best of our knowledge, so far this careful consideration has never been carried out and can be reproduced 
in many contexts where one needs to use coefficient-based model when  only 
finite number of non-equispaced observations are available.

Since our construction is based on application of Laguerre functions and the inversion of a triangular system, 
it  leads to very fast computations and produces no boundary effects that are due to the extension at zero and cut-off at $T$.
The presence of a tuning parameter $a$ allows for the choice of  the most advantageous time units, so
that both the kernel and the unknown right hand side of the equation are efficiently represented for
the further deconvolution.

 The methodology is illustrated by an extensive simulation study 
 using both earlier examples studied  in Abramovich {\it et al.}  (2013) and   
new settings based on the  kernels $g$ observed in the real DCE experiments.
Simulation study confirms that the proposed technique is fast, efficient, accurate,
practically usable and highly competitive: the new methodology  easily outperforms
the SVD, the Tikhonov regularization  and the kernel-based technique of Abramovich {\it et al.}  (2013).
The software is available on request for non-profit research purposes 
from Dr.  Yves Rozenholc ({\em yves.rozenholc@parisdescartes.fr}).

 %%%%%%%%%%%%%%%%%%%%%%%%%%%%%%%%%%%%%%%%%%%%%%%%%%%%%%%%%%%%%%%%%%%%%%%%%%%%%%%%%%%%%%%%%%%%

The rest of the paper is organized as follows. In Section~\ref{sec:Laguerre_functions}
we derive a system of equations resulting from expansion of the functions over the Laguerre
basis, study the effect of discrete, possible irregularly spaced data and introduce
selection of model size via penalization. Corollary~\ref{cor:risk_bound}
indeed confirms that the risk of the penalized estimator lies within a logarithmic factor of the
minimal risk. In Section~\ref{sec:asymp_risk} we extend our study  to the case $T\rightarrow\infty$ and 
provide asymptotic upper bounds for the risk proving that  the risk lies within 
a logarithmic factor of an oracle risk. The proof of this fact rests on
nontrivial facts of the theory of Toeplitz matrices. Section \ref{sec:motivation} considers 
high  frequency perfusion imaging as an important motivating example for the theoretical investigations of the paper.
Section~\ref{sec:simulation} provides an extensive 
simulation study. Section \ref{sec:real_data} presents an example of application of the methodology
developed in the paper to analysis of a DCE-MRI sequence of images of a participant 
of the REMISCAN cohort study \cite{remiscan} who underwent anti-angiogenic treatment for  renal cancer. 
Finally, Section~\ref{sec:discussion}
concludes the paper with discussion of results.  Section~\ref{sec:proofs}
contains some essential proofs. The rest of the proofs and other supplementary materials such 
as  introduction to theory of banded Toeplitz matrices and some of simulation results can be found 
in Section~\ref{sec:appendix},  Appendix.

\section{Laplace deconvolution via expansion over Laguerre functions basis }
\label{sec:Laguerre_functions}
\setcounter{equation}{0}

\subsection{Notations}

In what follows, we use letters  $f$, $g$ and $q$ for  functions $f(x)$, $g(x)$ and $q(x)$, respectively.
Vectors of values of those functions at points $t_1, \dots, t_n$ are denoted by $\vec f$, $\vec g$ and $\vec q$.
Vectors  of Laguerre coefficients are denoted by bold letters (e.g. $\mathbf{f_m}, \mathbf{g_m},
\mathbf{q_m}$) with the subscript indicating dimension of the vector. The coordinates of these vectors
are denoted with using superscripts: $\mathbf{f_m}=(f^{(0)}, \dots, f^{(m-1)})^T$, where $u^T$ denotes the transpose of $u$.

Given a matrix $\mathbf A$, let  $\mathbf A^T$ be the transpose of $\mathbf A$,
$\|\bA\|_2^2 = \Tr(\bA^T \bA)$ and $\rho^2(\bA) = \lambda_{\max} (\bA^T \bA)=\lambda_{\max} (\bA \bA^T)$ be, respectively,
the Frobenius and the spectral norm  of a matrix $\bA$, where $\lambda_{\max} (\mathbf U)$ is the largest,
 in absolute value, eigenvalue of $\mathbf U$. We denote by $[\mathbf A]_m$ the upper left $m\times m$
sub-matrix of $\mathbf A$. Given a vector $u \in {\mathbb R}^k$, we denote by $\| u \|$ its Euclidean
norm and, for $p\leq k$, the $p\times 1$ vector with the first $p$ coordinates of $u$, by $[u]_p$. For any
 function $t \in L_2(\mathbb R_+)$, we denote by $\| t \|_2$ its $L_2$ norm on $\mathbb R_+$.

\subsection{Coefficients of the Laguerre expansion and their estimators}
\label{sec:linear_system}
\label{sec:discrete_data}

In what follows, we assume that $f$ is square integrable over the positive half line ${\mathbb R}^+$.
Then, a common solution to the problem \fr{eq:model} is to represent $f$, $g$, $q$ and $y$ in
equations \fr{eq:model} and \fr{eq:Volterra} via some orthonormal basis on ${\mathbb R}^+$,
thus, reducing \fr{eq:model} and \fr{eq:Volterra} to a linear system of equations. 
It turns out that 
the Laguerre functions 
\be \label{eq:Laguerre_func}
\phi_k(t) = \sqrt{2a} e^{-at} L_k(2at),\ \ k=0,1, \ldots,
\ee
where $L_k(t)$ are Laguerre polynomials (see, e.g., Gradshtein and Ryzhik (1980))
$$
L_k(t) = \sum_{j=0}^k (-1)^j {k \choose j} \frac{t^j}{j!},\ \ \ t \geq 0,
$$
form a basis, which is particularly suitable for the problem at hand since 
it acts as a surrogate eigenfunction  basis for the problem (see, e.g. Weeks (1966) or Lien {\it et al.} (2008)).
Traditionally, one uses $a=1/2$, however, introduction of an additional parameter $a$ allows 
to choose the most appropriate time scale in the real-life applications of the methodology in general, and 
to perfusion imaging that motivates our study, in particular.

We denote by $f^{(k)}$, $g^{(k)}$, $q^{(k)}$ and $y^{(k)}$, $k = 0, \ldots, \infty$, the coefficients of the expansions over the Laguerre function basis of the functions $f(\cdot)$, $g(\cdot)$, $q(\cdot)$ and $y(\cdot)$ respectively.
By plugging these expansions into formula \fr{eq:Volterra}, we obtain the following
equation
\be \label{eq:laguerre1}
\sum_{k=0}^\infty q^{(k)} \phi_k(t) = \sum_{k=0}^\infty \sum_{j=0}^\infty f^{(k)} g^{(j)} \int_0^t \phi_k(x)  \phi_j (t-x) dx.
\ee
Due to the following relation (see, e.g., 7.411.4 in Gradshtein and Ryzhik (1980))
$$
\int_0^t \phi_k(x)  \phi_j(t-x)  dx = 2a e^{-at} \int_0^t L_k(2ax) L_j(2a(t-x)) dx =
(2a)^{-1/2} \,  [\phi_{k+j}(t) - \phi_{k+j+1} (t)],
$$
equation \fr{eq:laguerre1} can be re-written as
$$
\sum_{k=0}^\infty q^{(k)} \phi_k(t) = \sum_{k=0}^\infty \phi_k(t) [(2a)^{-1/2} \, f^{(k)} g^{(0)}  +
\sum_{\ell=0}^{k-1} (2a)^{-1/2} \, (g^{(k-\ell)}  - g^{(k-\ell-1)}) f^{(\ell)}].
$$
Equating coefficients for each of the basis functions, we obtain an infinite triangular 
system of linear equations. In order to use this system for estimating $f$, we denote 
the approximation of $f$ based on the first $m$ Laguerre functions by
\be \label{eq:fm}
f_m (x) = \sum_{k=0}^{m-1}  {f}^{(k)} \phi_k (x).
\ee
The following Lemma states how the coefficients in \fr{eq:fm} can be recovered.

\begin{lemma} \label{lem:triangular_system}
 Let $\bof_m$
%, $\bg_m$ 
and $\bq_m$ be $m$-dimensional vectors with elements 
$f^{(k)}$
%, $g^{(k)}$ 
and $q^{(k)}$, $k=0,1, \ldots, m-1$, 
respectively. Then, for any $m$, one has $\bq_m = \bG_m  \bof_m$ where
$\bG_m$ is the lower triangular Toeplitz matrix  with the first column 
$(g^{(0)},g^{(1)}-g^{(0)},\ldots,g^{(m-1)}-g^{(m-2)})^T\big / \sqrt{2a}$.
\end{lemma}

\noindent
Applying Lemma \ref{lem:triangular_system}  for $1\leq m\leq M$, we construct 
the following collection of estimators of $f(x)$ 
\be \label{eq:f_estimator}
\hat{f}_m (x) = \sum_{k=0}^{m-1} \hat{f}^{(k)} \phi_k (x)
\ee
where  $\widehat{\bof}_m  = (\hat{f}^{(0)}, \cdots, \hat{f}^{(m-1)}) = [\widehat{{\mathbf f_M}}]_m$.
Here,  $\widehat{{\mathbf f_M}} =\bG_M ^{-1} \widehat{\bq_M}$ and  
 \begin{equation}\label{eq:qMhat}
\widehat{{\mathbf q_M}}:=(\Phi_M^T\;\Phi_M)^{-1}\Phi_M^T\vec y
\end{equation}
is the unbiased  estimator of $(\Phi_M^T\;\Phi_M)^{-1}\Phi_M^T\vec q$
with $\vec q:=\left(q(t_1), \dots q(t_n)\right)^T$, $ \vec y:=(y(t_1), \dots, y(t_n))^T$ and
$$
\Phi_M:=\left(\begin{array}{ccc} \phi_0(t_1) & \dots & \phi_{M-1}(t_1) \\
\phi_0(t_2) & \dots & \phi_{M-1}(t_2)  \\ \vdots & \dots & \vdots  \\
\phi_0(t_n) & \dots & \phi_{M-1}(t_n)  \end{array}\right).
$$
Denoting $\bJ_{m,M}=\left( \bId_m \quad \mathbf 0_{m,M-m}\right)$  the $m\times M$
matrix which has the $m \times m$ identity matrix
$\bId_m$ as its first $m$ columns and the rest of the columns are equal to zero, 
the following relations hold
\begin{equation}\label{def:estim}
\widehat{{\mathbf f_m}}=[\widehat{{\mathbf f_M}}]_m=
[\mathbf G_M^{-1}(\Phi_M^T\;\Phi_M)^{-1}\Phi_M^T\vec y]_m=\mathbf G_m^{-1} \bJ_{m,M} \widehat{{\mathbf q_M}}=
\mathbf G_m^{-1}\bJ_{m,M} (\Phi_M^T\;\Phi_M)^{-1}\Phi_M^T\vec y.
\end{equation}
Note that, by using estimator (\ref{def:estim}) instead of the seemingly intuitive estimator
$\mathbf G_m^{-1} \widehat{{\mathbf q_m}}$, we manage to achieve two goals: 
avoiding re-fitting of the models for each value of $m$ and 
reducing the bias  that is due to having observations of the values of $\vec y$ 
rather than the noisy versions of Laguerre coefficients.

In order to understand the nature of this additional bias, observe that equation \fr{eq:model} is equivalent to 
\begin{equation}\label{eq:yinfty} 
\vec y=\Phi_\infty {\mathbf q_\infty}+ \sigma \vec \varepsilon 
\end{equation}
where $\vec \varepsilon:=\left(\varepsilon_1,  \dots,  \varepsilon_n\right)^T$ and  
$\Phi_\infty$ and $\mathbf{q_\infty}$ are the infinite versions of $\Phi_M$ and $\mathbf{q_M}$.
Consider vector   $\vec q_M:=\left(q_M(t_1), \dots, q_M(t_n)\right)^T$,
where $q_M(\cdot)$ is the orthogonal projection of $q$ on the  space spanned by the functions 
$\phi_0, \dots, \phi_{M-1}$. Then ,   $\vec q_M= \Phi_M {\mathbf q_M}$.
Heuristically replacing  $\Phi_\infty {\mathbf q_\infty}$ in (\ref{eq:yinfty})  by $\Phi_M {\mathbf q_M}$ and 
following the construction of  the linear regression  estimator, we estimate ${\mathbf q_M}$ by 
$\widehat{{\mathbf q_M}} =(\Phi_M^T\;\Phi_M)^{-1}\Phi_M^T\vec y$ as given by (\ref{eq:qMhat}).
Note that  ${\mathbf q_M}= (\Phi_M^T\;\Phi_M)^{-1}\Phi_M^T \vec q_M$ but 
${\mathbb E}(\widehat{{\mathbf q_M}})=(\Phi_M^T\;\Phi_M)^{-1}\Phi_M^T\Phi_\infty {\mathbf q_\infty}$, 
so   estimator $\widehat{{\mathbf q_M}}$ contains an additional bias ${\mathbf q_M} - {\mathbb E}(\widehat{{\mathbf q_M}})$ 
which we shall study later.

\subsection{The risk of the estimator}
\label{sec:model_selec}

We compute the mean integrated squared error (MISE):
\begin{eqnarray}
\nonumber {\mathbb E}( \|\hat f_m-f\|^2_2)&=& \|f-f_m\|_2^2+ {\mathbb E}
(\|\hat f_m-f_m\|^2_2) \\ \nonumber &= & \|f-f_m\|_2^2+ {\mathbb E}(\|\widehat{{\mathbf f_m}}-
{\mathbf f_m} \|^2)\\ \label{eq:dec1} & = &
\|f-f_m\|_2^2+  {\mathbb E}(\|\widehat{{\mathbf f_m}}-{\mathbb E}(\widehat{{\mathbf f_m}})\|^2)
+  \|{\mathbb E}(\widehat{{\mathbf f_m}})-{\mathbf f_m}\|^2
\end{eqnarray}
The first term in  the right-hand side of  \fr{eq:dec1}  is the functional  approximation bias resulting from replacing $f$ by its expansion over 
the finite system of Laguerre functions $(\phi_0, \dots, \phi_{m-1})$.
The second term is the variance term. The last term represents the additional 
bias which is due to estimation of the  coefficients in the orthonormal basis, defined on  the positive real line,
 using  a finite number of  data points that are  sampled on a finite interval   $[0,T]$. 
In order to control this last term, we introduce the following assumption\renewcommand{\thefootnote}{\textdagger}
\footnote{
Assumption {\bf (A0)} requires  $q$ defined by (\ref{eq:Fredholm}) to be smooth and decline as $t\rightarrow +\infty$.
More precisely, if $a=1/2$ for simplicity, it is sufficient that ${\mathcal L}^2 q \in {\mathbb L}^2({\mathbb R}^+)$
where the differential operator ${\mathcal L}$ is defined as
$$
{\mathcal L} u=-\left[t\frac{d^2}{dt^2} + \frac{d}{dt} -\frac t4\right]u.
$$
For details on Sobolev spaces associated to Laguerre functions, see Bongioanni and Torrea (2009), or Vareschi~(2015).
}
:\\

\noindent
 {\bf (A0):}\ \  For some   $C_q >0$, one has
$\displaystyle 
\sum_{k\geq 0} [k^2 q^{(k)}]^2 \leq C_q < \infty.
$
\\

\noindent
Denote   
$$
{\mathbf Q_m} =   \frac n T [(\Phi_M^T\Phi_M)^{-1}]_m ([\bG_M\bG_M^T]_m)^{-1}.
$$ 
Then, the following statement is true. 

\begin{proposition}\label{MISE}
Set  $M = M(n) = n^{(1+\eta)/3}$ where $0 < \eta <2$.   
If  Assumption {\bf (A0)} holds  and  $n \geq (2a C_q/ \sigma^2)^{1/\eta}$,   
then, 
\begin{equation}\label{eq:risk}
{\mathbb E}( \|\hat f_m-f\|^2_2)\leq \|f-f_m\|^2_2+
\frac 43 \frac{\sigma^2 T}{n} {\rm Tr}({\mathbf Q_m}) 
\end{equation}
and therefore
\begin{equation}\label{eq:riskbounds}
\|f-f_m\|^2_2+ \frac{\sigma^2T}n {\rm Tr}({\mathbf Q_m}) \leq
{\mathbb E}( \|\hat f_m-f\|^2_2)
\leq   \|f-f_m\|^2_2+ \frac 43\frac{\sigma^2T}n {\rm Tr}({\mathbf Q_m}) 
\end{equation}
\end{proposition}

\noindent
The proofs of this and the later statements are presented in Section~\ref{sec:proofs}, Proofs, 
or in Appendix~\ref{sec:appendix}.

\begin{remark} \label{rem:M}
{\rm   The choice of the value of $\eta$ in Proposition \ref{MISE} depends on how large the 
number of observations $n$ is.  The medium value $\eta = 1$ corresponds to the very moderate requirement  $n \geq 2a C_q/ \sigma^2$ on the value  of $n$.
If $n$ is relatively large, one can reduce $\eta$ and, therefore,  $M$ since smaller values of $M$ lead to more stable computations. For instance, if one selects $M(n) = n^{1/2}$, then $\eta = 1/2$ and the estimator is fully adaptive  as long as $n \geq (2a C_q/ \sigma^2)^2$.  }
\end{remark}

%\medskip

%%%%%%%%%%%%%%%%%%%%%%%%%%%%%%%%%%%%%%%%%%%%%%%%%%%%%%%%%%%%%%%%%%%%%%%%%%%%%%%%%%%%%%%%%%%%%%%%%%%%%

We define the set of indices
\begin{equation}\label{def:Mn}
\mathcal M_n   = \{1,\ldots,M\}.
\end{equation}
The smallest possible risk, the so-called {\it oracle } risk,   is obtained by minimizing the left-hand side
of expression \fr{eq:riskbounds} with respect to $m$:
\begin{equation}\label{eq:oracle}
R_{oracle}  =
\min_{m\in{\mathcal M}_n}\, \left[ \|f_m - f\|^2_2 + \sigma^2  T n^{-1} \ \Tr(\bQ_m)  \right].
\end{equation}

Hence, the objective is to choose a value of $m\in{\mathcal M}_n$ which delivers 
an  estimator of the unknown function $f$ with the risk as close  as possible to 
the oracle risk or at least to the right-hand side of (\ref{eq:oracle}). 
Since the bias term $\|f_m - f\|^2_2$ is unknown, in order to attain this goal, one can use
a penalized version of estimator \fr{def:estim} as it is described in the next section.

%%%%%%%%%%%%%%%%%%%%%%%%%%%%%%%%%%%%%%%%%%%%%%%%%%%%%%%%%%%%%%%%%%%%%%%%%%%%%%%%%%%%%5

\subsection{Selection of the model size via penalization}

Denote
 \begin{equation}\label{def:Am}
\mathbf A_m = \sqrt{\frac nT}\, \mathbf G_m^{-1}\bJ_{m,M} (\Phi_M^T\;\Phi_M)^{-1}\Phi_M^T.
 \end{equation}
and 
\be \label{eq:vm_rhom}
  v_m^2 := \|{\mathbf A}_m \|^2_2 = \Tr(\bQ_m),\ \ \ \rho_m^2 :=  \lambda_{\mathrm{max}}(\bA_m^T \bA_m)   
\ee
Introduce the penalty
\be \label{eq:penalty}
\pen(m) :=  8 \sigma^2 T n^{-1} \lkv v_m^2 + 2\, \kappa \rho_m^2  \, \log (m\,\rho_m / \rho_1) \rkv,
\ee
where $\kappa=1$ for Gaussian errors $\varepsilon_i$ and $\kappa$ is the squared sub-gaussian norm of  $\varepsilon_i$,  
otherwise (see the definition in Vershynin~(2012)).  
The value $\rho_1 = \bA_1^T \bA_1= \| \bA_1 \|^2$ is  the squared norm of vector $\bA_1$ and is necessary to account for the   
scale parameter $a$.

For each $m=1,\ldots,M$,  consider the estimator $\widehat{f}_m$ of $f$ of the form (\ref{eq:f_estimator}) where the coefficients 
$\widehat{\mathbf f_m}$ are defined by (\ref{def:estim}). This estimator appears as the least squares   estimator with the 
contrast equal to
$-\| \widehat{\mathbf f_m} \|^2$.
For selecting the model size $m$,  we search for  $\widehat{m}$ which minimizes the sum of the penalty and the contrast
\be \label{eq:hatm}
\widehat{m} := \arg \min \left\{m \in {\mathcal M}_n:\  -\| \widehat{\mathbf f_m} \|^2 + {\rm pen}(m) \right\}.
\ee
and obtain the penalized least squares estimator  $\widehat{\mathbf f_{\widehat m}}$ of the vector of   Laguerre coefficients.
Finally, we construct   the   estimator $\hat f_{\widehat m}$ of $f$ using Laguerre coefficients $\widehat{\mathbf f_{\widehat m}}$.\\

The heuristic argument behind this model selection procedure is the following. Since $\|f-f_m\|_2^2 = \|f\|_2^2-\|f_m\|_2^2$,
the bias-variance balance is attained by the value $\widehat{m}$ of  $m$ that delivers the minimum
of $-\|f_m\|_2^2+ \Var (\widehat f_m)$. The term $\|f_m\|_2^2$ is estimated
by $\|\widehat f_m\|_2^2=\|\widehat{\mathbf f}_m\|^2$ and the variance term is approximated by ${\rm pen}(m)$.  
Indeed, the  following statement  holds.

\begin{theorem} \label{th:risk_bound}
Let Assumption {\bf (A0)} hold  and  $n \geq (2a C_q/ \sigma^2)^{1/\eta}$. If  $M=M(n)= n^{(1+\eta)/3}$,  
then  one has
\be \label{eq:risk_bound}
\EE (\|\hat{f}_{\widehat{m}} - f\|^2_2) \leq  \min_{m \in {\mathcal M}_n}  \left[ 9 \|f_m - f\|_2^2 + 6 {\rm pen}(m)    +
72 \sigma^2 \rho_1^2 \, \frac{T}{mn} \right].
\ee
\end{theorem}

Since    $\rho_m^2 \leq v_m^2$  for any value of $m$, it follows from Theorem \ref{th:risk_bound}
that,  for any value of $m$, the risk of the estimator $\hat{f}_{\widehat{m}}$ lies within a
logarithmic factor of the upper bound of oracle risk defined in (\ref{eq:oracle}).
Note that the upper bound in Theorem \ref{th:risk_bound} is non-asymptotic and holds
for any values of $T$ and $n$ and any distribution of points $t_i$, $i=1, \ldots, n$.
In particular, the following corollary  is valid.

\begin{corollary} \label{cor:risk_bound}
Under conditions of Theorem \ref{th:risk_bound}, one has
\be \label{eq:risk2}
\EE (\|\hat{f}_{\widehat{m}} - f\|^2_2) \leq 48 [1+ 2\kappa\log (m_0\rho_{m_0}/\rho_1)] R_{oracle} +
96 \sigma^2 \rho_1^2 \frac T{m_0n},
\ee
where $m_0=m_0(n, T)$ is the value of $m$ delivering the minimum in the right-hand side of \fr{eq:risk_bound}.
\end{corollary}

%%%%%%%%%%%%%%%%%%%%%%%%%%%%%%%%%%%%%%%%%%%%%%%%%%%%%%%%%%%%%%%%%%%%%%%%%%%%%%%%%%%%%%%%%%

\section{Asymptotic upper bounds for the risk and optimality of the estimator }
\label{sec:asymp_risk}
\setcounter{equation}{0}

Corollary \ref{cor:risk_bound} is valid for any function $g$ and any distribution of sampling points,
hence, it is true in the ``worst case scenario''.  It does not allow one to judge how fast   
the risk decreases when $n$ grows. In particular, since the problem of Laplace deconvolution is 
an ill-posed problem, one needs to know how fast   the error grows when $m$ increases.
Abramovich {\it et al.} (2013)  addressed this question by showing that, under certain assumptions, the risk
of the kernel density estimator grows as a negative  power of the bandwidth, so that 
the overall error tends to zero  at a polynomial rate. In what follows, we introduce assumptions similar to those of 
Abramovich {\it et al.} (2013) and show that the MISE of the estimators produced by our methodology 
grows as a power of the model size, so that $\log(\varrho_m)$ is just a multiple of $\log m$. 
Moreover, we  establish that the spectral   and the Frobenius norms
of matrix ${\mathbf A}_m$ grow at the same rate as $m$ increases.

\subsection{Assumptions}

Let $r \geq 1$  be such that
\be \label{k_cond}
\left. \frac{d^j g(t)}{dt^j} \right|_{t=0} = \lfi
\begin{array}{ll}
0, & \mbox{if}\ \ j=0, ..., r-2,\\
B_r \ne 0, &  \mbox{if}\ \ j=r-1,
%\\ \ne 0
\end{array} \right.
\ee
with the obvious modification $g(0)=B_1 \ne 0$ for $r=1$.
Consider matrix
\begin{equation}\label{eq:bOm}
\bOm_m:= \frac nT [(\Phi_M^T\Phi_M)^{-1}]_m
\end{equation}
and assume that function $g(x)$, its Laplace transform $G(s):=\int_0^{+\infty} e^{-sx} g(x)dx$, 
and   matrix $\bOm_m$ satisfy the following conditions.

\begin{itemize}

\item[{\bf (A1)}]  $g \in L_1 [0, \infty)$ is $r$ times differentiable with   $g^{(r)} \in L_1 [0, \infty)$.

\item[{\bf (A2)}] Laplace transform $G(s)$ of $g$ has no zeros with nonnegative real parts except for zeros of the form $s=\infty + ib$.

\item[{\bf (A3)}]  There exists $n_0$ such that, for $n>n_0$, eigenvalues of matrix  $\bOm_m$  are uniformly
bounded, i.e.
\be  \label{eigenvalues}
% \forall m=1, \dots, M, \;\;
 0< \lambda_1 \leq \lambda_{\min}(\bOm_m) \leq \lambda_{\max}(\bOm_m) \leq \lambda_2 < \infty
\ee
for any $m=1, \dots, M,$ and  some absolute constants $\lambda_1$ and $\lambda_2$.

\end{itemize}

Consider, for example,   
\be \label{g123}
g_1(t)=e^{-5t} (2t-\sin(2t)),\qquad g_2(t)=e^{-5t},\qquad g_3(t)=e^{-t} (2t+1),
\ee
Then, 
$g_1 (t)=e^{-5t} (2t-\sin(2t))$ and   then $g_1(0) = g_1'(0) = g_1''(0)=0$ and $g_1'''(0) = 8$,
so that $r=4$ for $g_1$ and $r=1$ for $g_2$ and $g_3$. 
One can also easily evaluate Laplace transforms $G_1 (s) = 8 (s+5)^{-2}[(s+5)^2 +4]^{-1}$,
$G_2(s) = (s+5)^{-1}$ and $G_3(s) = (s+1)^{-2} (s+3)$. Hence, functions $G_1 (s)$ and $G_2 (s)$
do not have zeros and $G_3 (s)$ has a single zero $s_1= -3$ with a negative real part. 
Later, we shall use the kernels \fr{g123} in our simulation study.

Definition of $r$ and  Assumptions {\bf (A1)} and {\bf (A2)} are similar to those introduced in Abramovich {\it et al} (2013).
Assumption {\bf (A1)} requires $g(t)$ to have $r$ derivatives and to decline as $t \to \infty$
Assumption {\bf (A2)} establishes that the Toeplitz matrix does not have eigenvalues that decrease exponentially 
as the functions of the matrix dimension. Finally, Assumption {\bf (A3)}  ensures that 
the design points $t_i$, $i=1, \cdots, n,$ are relatively regularly spaced on the interval $[0, T]$.
The normalization of $\bOm_m$ by $T/n$ is justified by the fact that the matrix tends to the identity matrix 
when both $n$ and $T$ tend to infinity. Assumption {\bf (A3)} also implies that $M \leq n$.

Observe that, if $g$ is known exactly, all   assumptions are set on known quantities.
 If $g$ is  known only approximately (or is estimated from data 
as in the case of DCE imaging),  the value  of $r$ and the locations of zeros of $G(s)$
are hard to determine. However,  Assumption {\bf (A3)} is independent of $g$ and can always be verified. 
In particular,  one can compute matrices $\bOm_m$ and find their lowest and highest eigenvalues 
$\lambda_1$ and $\lambda_2$. Nevertheless, unlike in Abramovich {\it et al} (2013), our estimation technique does 
not rely on the knowledge of $r$ or $G(s)$, so that the risk satisfies the oracle inequalities 
\fr{eq:risk_bound} and \fr{eq:risk2} whether we know those quantities or not.

\subsection{Asymptotic near-optimality of the estimators}

From properties of Toeplitz matrices that are reviewed in the Appendix, it follows that
under Assumptions {\bf (A1)}--{\bf (A3)}, both $v_m^2$ and $\rho_m^2$ are polynomial in $m$.
Moreover, Lemma \ref{lem:rho_v}  presented in Appendix (Section \ref{sec:appendix})
shows that,  for $m$   large enough,  one has
\be \label{eq:vm_bounds}
C_1   m^{2r} \leq  \, \rho_m^2    \leq  v_m^2 \leq   C_2    m^{2r}, \quad
\ee
for some absolute positive constants $C_1$ and $C_2$, exact values of which are presented
in Lemma~\ref{lem:rho_v}. Hence,   Lemma~\ref{lem:rho_v} implies that, in   \fr{eq:penalty},
$\rho_m^2 \log(m\rho_m/\rho_1) \propto v_m^2\log(m)$   as  $m \rightarrow \infty$,   so that
the second term  in \fr{eq:penalty} is  almost of the same asymptotic order as the first   term, up to at most $(\log n)$ factor.
Consequently, as $n \rightarrow \infty$ and $T/n \rightarrow 0$,  the right-hand side of \fr{eq:risk_bound}
is of almost the same asymptotic order as  the oracle risk \fr{eq:oracle}.
Thus, by combination of Theorem \ref{th:risk_bound} and Lemma \ref{lem:rho_v}, we obtain to  the following statement.

\begin{theorem} \label{th:asymp_risk}
% Let  condition \fr{eq:rho_condition}  hold for some positive constants $\alpha$ and $C_\rho$.
Under assumptions {\bf (A0)}--{\bf (A3)}, for an estimator $\hat{f}_{\widehat{m}}$ of $f$ with penalty given by equation
\fr{eq:penalty},    as $n \rightarrow \infty$,
\be \label{eq:optimality}
\frac{R(\hat{f}_{\widehat{m}}) }{R_{oracle}} \leq  C(r) \log n\ [1 + o(1)],
%\ \frac{T}{n} \rightarrow 0,\ m \rightarrow \infty,\ \frac{m^{2r+1} T}{n} \rightarrow 0.
\ee
provided $T/n \rightarrow 0$  as $n \rightarrow \infty$. Here,  $C(r)$ is a constant that depends on $r$ only.
\end{theorem}

\begin{remark} \label{rem:T}
{\rm  The theory above is valid for $T$ being finite as well as for $T=T_n \rightarrow \infty$
as long as $T_n/n \rightarrow 0$ as  $n \rightarrow \infty$. Indeed, the natural consequence of $T$
being finite is that the bias term $\|f-f_m\|_{2}^2$ might be relatively large due to misrepresentation
of $f$ for $t>T$. However, since both the risk
of the estimator $R(\hat{f}_{\widehat{m}})$ and the  oracle risk are equally affected,
Theorem \ref{th:asymp_risk} remains valid whether $T=T_n$ grows with $n$ or not.}
\end{remark}

\section{Motivation: perfusion imaging and DCE imaging data}
\label{sec:motivation}

Cancers  and vascular diseases inducing stroke and heart infraction  present major public health concerns. 
Considerable improvement in assessing the quality of a vascular network and its permeability have 
been achieved through perfusion imaging using Dynamical Contrast Enhanced (DCE) imaging procedures 
with either Computer Tomography (DCE-CT),  Magnetic Resonance Imaging  (DCE-MRI) or Ultra Sound (DCE-US).
The common feature of DCE imaging techniques is that each of them uses the rapid injection of a single dose of a bolus of a contrast agent  
and monitors its progression in the vascular network by sequential  imaging.
Currently, the high frequency DCE imaging techniques are  more and more commonly   used for medical assessment  
of  brain flows for prognostic and therapeutic purposes after stroke, or, of  
cancer angiogenesis. They  have a great potential for cancer detection and
characterization, as well as for monitoring \textit{in vivo} the effects of treatments
(see, e.g., Cao (2011); Cao {\it et al.}  (2010); Goh {\it et al.}  (2005); Goh and Padhani  (2007);
Cuenod {\it et al.}  (2006); Cuenod {\it et al.}  (2011);  Miles  (2003);  Padhani and Harvey   (2005) and Bisdas
{\it et al.}  (2007)).

\begin{figure}
%\[ \includegraphics[height=6cm]{blood-net7.pdf}\]
 \[ \includegraphics[height=6cm]{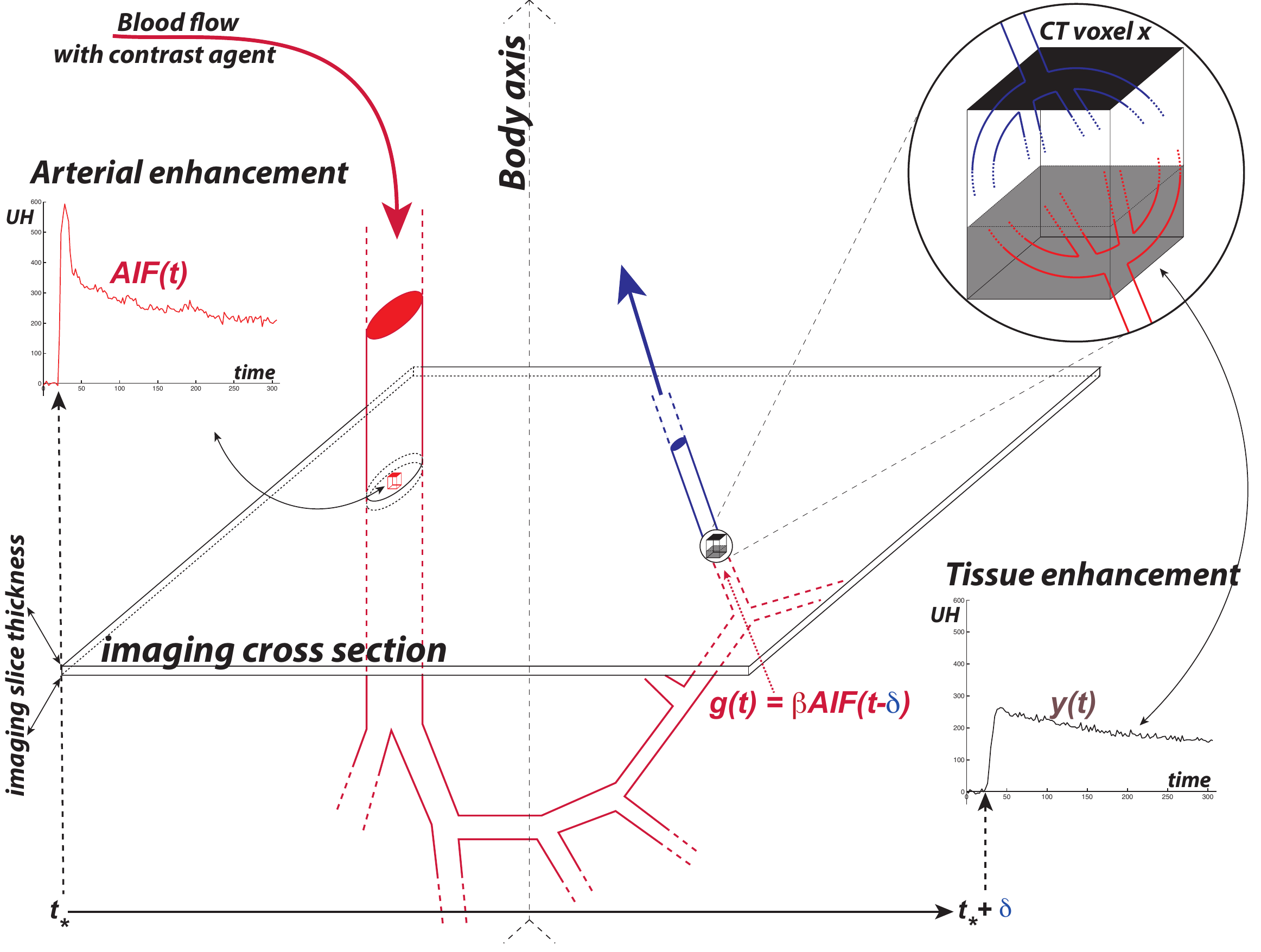}\]
\caption{\small \label{fig:DCE-CT}  {\bf DCE imaging experiment and contrast agent circulation.}
Figure shows a sub-tree of the vascular system going from the artery --which receives   oxygenated blood 
(red arrow)-- to the vein  --which returns the de-oxygenated blood (blue arrow) after exchanges within the tissue. 
After passing through the heart, the bolus of the contrast agent, injected  into a vein,  is distributed,  throughout the body 
along the arterial network to the tissue and later back to the venous system. In the imaging cross-section, the contrast 
agent induces enhancements first in the artery, providing the AIF, and later in the tissue of interest providing   
observations $y(t_i)$, $i=1,\ldots, n$. Enhancements are measured in the voxels of the imaging cross-section.} 
\end{figure}

  As an example, below we consider a DCE experiment which  follows propagation, through the vascular network,  of a bolus of a contrast agent, 
injected in a vein, after it passes through the heart.   Assuming that all voxels have unit volumes, at a microscopic level,
for a given tissue voxel of interest, the number of arriving particles at time $t+\delta$ is given by
$\beta\, \AIF(t)$. Here, $\AIF(t)$ is the Arterial Input Function that measures concentration of  the contrast agent within  
 the tissue voxel inside the aorta at time $t$, and parameter $\beta$, the  so-called \textit{Tissue Blood Flow},
 is the  proportion of the contrast agent  which enters this  voxel. 
Denote the number of particles in the tissue voxel   at time $t$ by $y(t)$ and the random lapse of time
during which a particle sojourns in the tissue voxel by $S$. Assuming sojourn times for different particles
to be independent and identically distributed  with  a cumulative distribution function  $F$, 
one obtains the following equation for the average number of
 particles of the contrast agent in the tissue voxel at the moment $t$
$$
{\bb E} y(t) =  \hspace{-3mm}\underbrace{ \int_0^{t-\delta} \beta\, \AIF (t-\tau)\,d\tau}_{\text{arrived before time $t$}}
\hspace{-1mm}- \underbrace{ \int_0^{t-\delta}  \beta\, \AIF (t-\tau)\,P(S \leq \tau)\,d\tau}_{\text{left before time $t$}}
= \int_0^{t-\delta}  \, \AIF (t-\tau)\,\beta(1-F(\tau)) d\tau,
$$
where the expectation is taken under the unknown distribution of the sojourn times and   $\delta$ is the
delay  between the measurement of the concentration of the contrast agent  inside the aorta % (first pass inside the imaging section)
 and its arrival inside the tissue voxel of interest. % (second pass inside the imaging section). 
Assuming that the transit inside the arteries is homogeneous,  up to   parameter $\delta$, 
the aorta acts as a good proxy of the feeding artery of the voxel of interest. 
In reality, one does not know ${\bb E} y(t) $ and has discrete noisy observations
$$
y(t_i) = {\bb E} y(t_i) + \sigma \varepsilon_i,
$$
where $\varepsilon_i$ are i.i.d. standardized random variables.

Medical doctors are interested in a reproducible quantification of the blood flow inside
the tissue which is characterized by $f(t)= \beta(1-F(t))$
since this quantity is independent of the concentration of particles of contrast agent
within a  voxel inside the aorta  described by $\AIF(t)$.  
% Here, we should notice that $f(t)$ is not a 
% dimensionless quantity as the tissue blood flow, $\beta$, is expressed in $mL/min/100mL$ 
%  ($mL$ of blood passing through 100 $mL$ of tissue within one minute). 
The sequential imaging acquisition is illustrated by Figure \ref{fig:DCE-CT}.
The contrast agent arrives with the oxygenated blood through the aorta (red arrow)
where its concentration, AIF, within unit volume voxel is measured first  when it passes
through the imaging cross-section (red box).
Subsequently, the contrast agent enters the arterial system, and it is assumed
that its concentration does not change during this phase. The exchange within the tissue
of both oxygen and contrast agent occurs  from the beginning of the feeding phase  and the concentration
of contrast agent during this exchange is measured in all tissue voxels
inside the imaging cross-section (grey voxel in the zoom). Later the contrast agent returns to the venous system with the
de-oxygenated blood (blue arrow).

This leads to the following complete observation model:
\beqn \label{eq:AIF model}
y(t_i) & = & \int_0^{t_i-\delta} \, \AIF(t_i - \tau)\  \beta (1-F(\tau)) d\tau + \sigma \varepsilon_i,  \quad i=1,...,n,\\ 
\eta(t_i) & = & \AIF(t_i) + \sigma_0 \xi_i, \quad j=1,...,m,
\eeqn
where $\xi_i$,    are i.i.d. centered random variables independent from the $\varepsilon_j$, $j=1, \cdots, n$. 
The value of delay $\delta$ can be  measured with a small error using the delay between
the moment when the contrast agent appears inside the aorta and the time it appears
in the voxel of interest -- both being measured in the imaging cross section. 
For this reason, in what follows, we assume that the time measurements are
appropriately shifted, so that we can use $\delta=0$ in \fr{eq:AIF model}.
Unfortunately, evaluation of the proportion $\beta$ is a much harder task and, hence,  is realized  with
a much larger error. Mathematically, it corresponds to estimation of the value of $f$ at $t=0$ since  $F(0)$ is always zero.

In addition, a large artery, like the aorta, when available in the imaging field, usually covers a Region Of Interest (ROI) of few hundreds voxels.
In this case,  the observed value $\eta(t_i)$ is obtained by averaging (at each time $t_i$) of the values observed in the ROI leading to 
$\sigma_0 \ll \sigma$, so that we can assume that $\sigma_0=0$. Therefore, the complete model \fr{eq:AIF model}  for DCE imaging experiments reduces to
the Laplace convolution equation based on noisy observations of the form \fr{eq:model}, the study of which  
presents  a necessary theoretical step before obtaining medical answers on the basis  of the model \fr{eq:AIF model}.
Nevertheless, we draw attention to the fact that, in the DCE context, 
  $\AIF(t)$ is only available at the observation times $t_i$, $i=1,\ldots,n$.

%%%%%%%%%%%%%%%%%%%%%%%%%%%%%%%%%%%%%%%%%%%%%%%%%%%%%%%%%%%%%%%%%%%%%%%%%%%%%%%%%%%%%%%%%%%%%%%%%%%

\section{Simulation study}\label{sec:simulation}
 
In this section we present the results of a simulation study to illustrate finite sample performance of the Laplace deconvolution 
procedure developed above. In what follows, we compare our method with the one introduced in Abramovich {\it et al.} (2013), since, 
to the best of our knowledge, it is the only competitive method specifically designed for solution 
of  Laplace  convolution equation in the presence of noise. We also carried out  comparisons with the 
standard techniques designed for  solution of general ill-posed linear inverse problems, namely, the
Tikhonov regularization and the Singular Value Decomposition (tSVD).

Moreover, we put our best effort to apply  the Laplace transform inversion of the numerical 
realization of the Laplace transform of our equation suggested by Mnatsakanov  (2011)
and Mnatsakanov  and Sarkisian  (2013) but failed to produce any reasonable results due 
to the small sample sizes ($n\leq 250$).

\subsection*{Settings}

We used two different simulation settings. In the first one, an analytic 
form of $g$ is known, so that  the estimator $\hfapr$ developed in Abramovich {\it et al.} (2013) is available. 
In the second setting, only $g(t_1)$, \ldots, $g(t_n)$ are known, so that one cannot construct $\hfapr$. 
In both setting, we  considered Gaussian noise in \fr{eq:model} and set $\kappa=1$ in \fr{eq:penalty}.

\paragraph{Setting 1:  $g$ exactly known. }  We use the simulation set up of Abramovich {\it et al.} (2013). 
In particular, we considered fixed regular design with $T=10$, sample sizes $n=100$ and $250$,
 and three choices of the true function: 
$f_1(t)=t^2 e^{-t}$, $f_2(t) = 1-\Gamma_{2,2}(t)$ and $f_3(t) = 1-\Gamma_{3,0.75}(t)$, where
$\Gamma_{\alpha,\theta}$ is the c.d.f of the Gamma distribution with the shape parameter $\alpha$ and the scale parameter $\theta$. 
We used the five convolution kernels $g_1$, \ldots, $g_5$,  where $g_1$, $g_2$ and $g_3$ are defined in \fr{g123}
% \be \label{g123}
% g_1(t)=e^{-5t} (2t-\sin(2t)),\qquad g_2(t)=e^{-5t},\qquad g_3(t)=e^{-t} (2t+1),
% \ee
and kernels $g_4$ and $g_5$ are of the forms
$$
g(t) =   e^{-3t} t^{2} \sum_{j=0}^k \frac{\rho_j}{(j+2)!} t^j,
\quad \mbox{with} \quad
G(s) = (s+3)^{-(k+3)}\, \sum_{j=0}^k \rho_j (s+3)^{k-j},  %\ \ \rho_0=1.
$$
their Laplace transforms.
Here, $\rho_0=1$;   $k=4$ and the numerator of $G(s)$ has four roots $(-4 \pm 2.5i,  -0.75\pm 1.5i)$
for $g_4$;   $k=6$ and the numerator of $G(s)$ has six roots $(-4 \pm 2.5i,  -0.75\pm 1.5i,  -2 \pm 2i)$ 
for $g_5$. Both $g_4$ and $g_5$ are such that $r=3$ in (\ref{k_cond}). For each kernel, we chose the nominal noise levels $\sigma_0(g_j)$ that were, respectively,  equal to 
0.001, 0.1, 0.01, 0.002, 0.002 for $g_1, \ldots, g_5$. 
Simulations were carried out with noise levels $\sigma_0(g_j)/2^i$, $i=1,\ldots, 5$.

\begin{figure}
%\[ \includegraphics[height=4cm]{real-kernels.pdf}\]
 \[ \includegraphics[height=4cm]{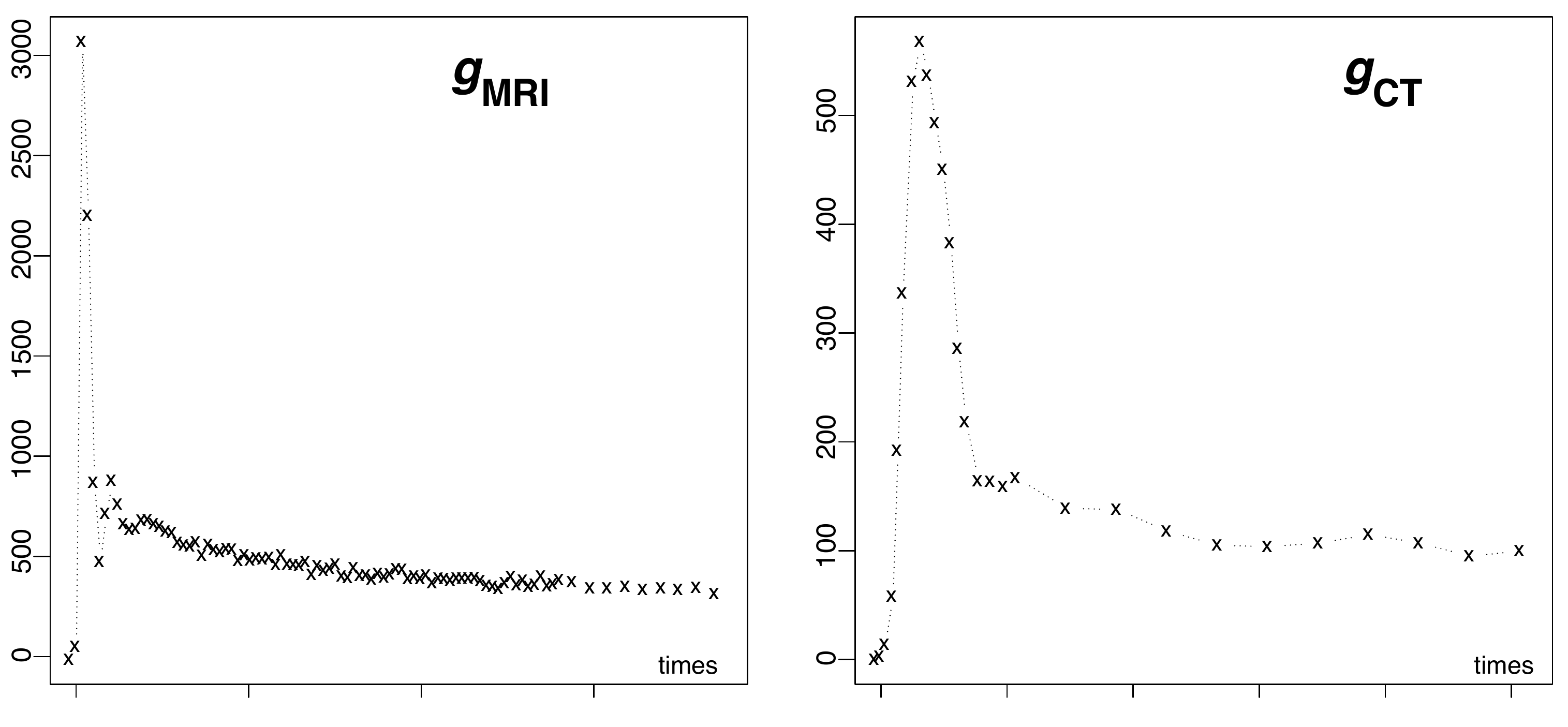}\]
\caption{\small \label{fig: real kernels}  {\bf DCE kernels:} 
(left) from a DCE-CT experiment ($n=91$) and (right) from a DCE-MRI experiment ($n=28$). 
Baselines are removed to access only the enhancements produce by the arrival of the bolus of the contrast agent. 
Crosses correspond to the observation times that are rescaled to the interval $[0,10]$.} 
\end{figure}

\paragraph{Setting 2: observations of  $g$ are available.} We consider 
two  ``real life" kernels,  $g_{\mathrm{MRI}}$ and $g_{\mathrm{CT}}$, obtained, respectively, from a DCE-MRI ($n=91$) and a DCE-CT ($n=28$)
 sequences of one patient. For those kernels, shown in Figure \ref{fig: real kernels}, only $g(t_i)$, $i=1,\ldots,n$ are observed 
and no analytical form of $g$ is available.  In (\ref{eq:AIF model}), we chose $\beta = 0.5$ 
and $f_2$, $f_3$ and  $f_4(t)=\exp(-2\,t)$ as test functions, since they correspond to typical survival functions $1-F$. 
For each test function, we computed the function $q$ at the time points $t_i$, $i=1,\ldots,n,$ in \fr{eq:Volterra} 
by numerical integration with trapezoid rule. Then we added Gaussian noise  with realistic noise levels, namely,
 $\sigma=60$ for $g_{\mathrm{MRI}}$ and $\sigma=25$ for $g_{\mathrm{CT}}$. \\

\subsection*{Implementation}

In order to implement our procedure and to stay as close as possible to the real-life DCE imaging  experiments, 
we only used the knowledge of the vector $\vec g=(g(t_1),\ldots,g(t_n))^T$ of values of $g(t)$ at the points $t_1, \cdots, t_n$.
The elements of  matrix $\bG_M$ are  derived from $\widehat{{\mathbf g_M}}$, the linear regression estimators 
of the Laguerre coefficients of $g$,  obtained as $\widehat{{\mathbf g_M}}  =(\Phi_M^T\;\Phi_M)^{-1}\Phi_M^T\vec g$,
similarly to   \fr{eq:qMhat}.

We implement our procedure using the public software $R$. Numerical computations using Laplace functions 
are facing numerical instabilities when $M$ is too large. Hence, for a given kernel $g$ and  for each value $a$, 
we selected the largest value of $M \leq 25$ such that both matrices 
$\Phi_M^T\Phi_M$ and $\widehat{\bG_M}\widehat{\bG_M}^T$ are of full rank and set $M(n) =M$  in \fr{def:Mn}  and 
Proposition \ref{MISE}. For the sample sizes $n=100, 250, 91$ and  28, used in our simulation settings, this leads to 
$\eta=1.10, 0.75, 1.14$ and $1.90$, respectively in Proposition \ref{MISE} and  Theorem \ref{th:risk_bound}.

Subsequently, we derived $\widehat m$ using \fr{eq:hatm} and obtained the penalized  estimator $\widehat{\mathbf f_{\widehat m}}$ 
of the vector of Laguerre coefficients.  We evaluated the estimator $\widehat{q} = \widehat{q} (a)$ 
on the basis of the estimator $\widehat{\mathbf f_{\widehat m}}$. At last, we chose the value of $a$ 
which minimizes the Euclidean norm  $R(a)$ of the difference between $\vec y$ and the vector $\widehat{q} (a)$.

\subsection*{Competing techniques}

We compared our procedure (referred to as $\widehat f_{\mathrm{LAG}}$ with the estimator introduced in Abramovich {\it et al.} (2013), 
(denoted $\widehat f_{\mathrm{APR}}$ below) as well as with the Tikhonov regularization ($\widehat f_{\mathrm{TIKH}}$ ) and  
the Singular Value Decomposition ($\widehat f_{\mathrm{tSVD}}$). 
To this end, we rewrote equation \fr{eq:model} using   trapezoidal approximation of the integral $\int_0^t g(t - \tau) f(\tau) d\tau$,
thus, realizing  the Laplace convolution as {$\vec q \approx A \vec f$} %$\vec y \approx A \vec f$ 
where $A$ is the lower triangular matrix.
We considered the SVD  of $A$, $A=U\, S \,V^T$ where, respectively, $S$ is a diagonal and $U$ and $V$
are orthogonal matrices. Then, the Tikhonov regularization -based estimator % (denoted by $\widehat f_{\mathrm{TIKH}}$)
is given by
\be \label{Tikhonov}
\widehat f_{\mathrm{TIKH}} = V\,  S\, (S^2 + \lambda I)^{-1}\, U^T \, U^T\,\vec y
\ee
The   SVD estimator  is defined as  %(denoted by $\widehat f_{\mathrm{tSVD}}$)
\be \label{eq:SVD}
\widehat f_{\mathrm{tSVD}} = V \,   (S_k)^{-1}  \, U^T\,\vec y,
\ee
where $S_k$ is the diagonal matrix derived  from $S$ by setting its $k$ smallest components to infinity,
so they vanish in $(S_k)^{-1}$.
The values of parameters $\lambda$ in \fr{Tikhonov} and $k$ in \fr{eq:SVD} are obtained 
 by minimizing the Euclidean norm of the difference between $\widehat q$ (reconstructed from, respectively, 
$\widehat f_{\mathrm{TIKH}}$ and  $\widehat f_{\mathrm{tSVD}}$), and an estimated version $\tilde{q}$ of $q$ 
obtained from $y$  by local polynomial regression fitting.

\subsection*{Results of simulations}

 For each simulation, given an estimator $\widehat f$ of $f$ at times $0=t_1\leq\ldots\leq t_n=T$, 
in order to take into  account the possibly irregular design, we computed the Integrated Square Errors, 
$ISE(\widehat f)$ over the interval $[0,T]$ using the trapezoidal approximation of the integral. 
 In each   setting, we carried out 400 simulation runs. For each estimator, we calculated the average 
values of $ISE(\widehat f)$ over those runs and the corresponding standard deviations.

\begin{figure}
%\[ \includegraphics[width=12cm]{BoxPlotRisksRatio-full-n100-P2-c2.pdf}\vspace{-8mm}\]
\[ \includegraphics[width=12cm]{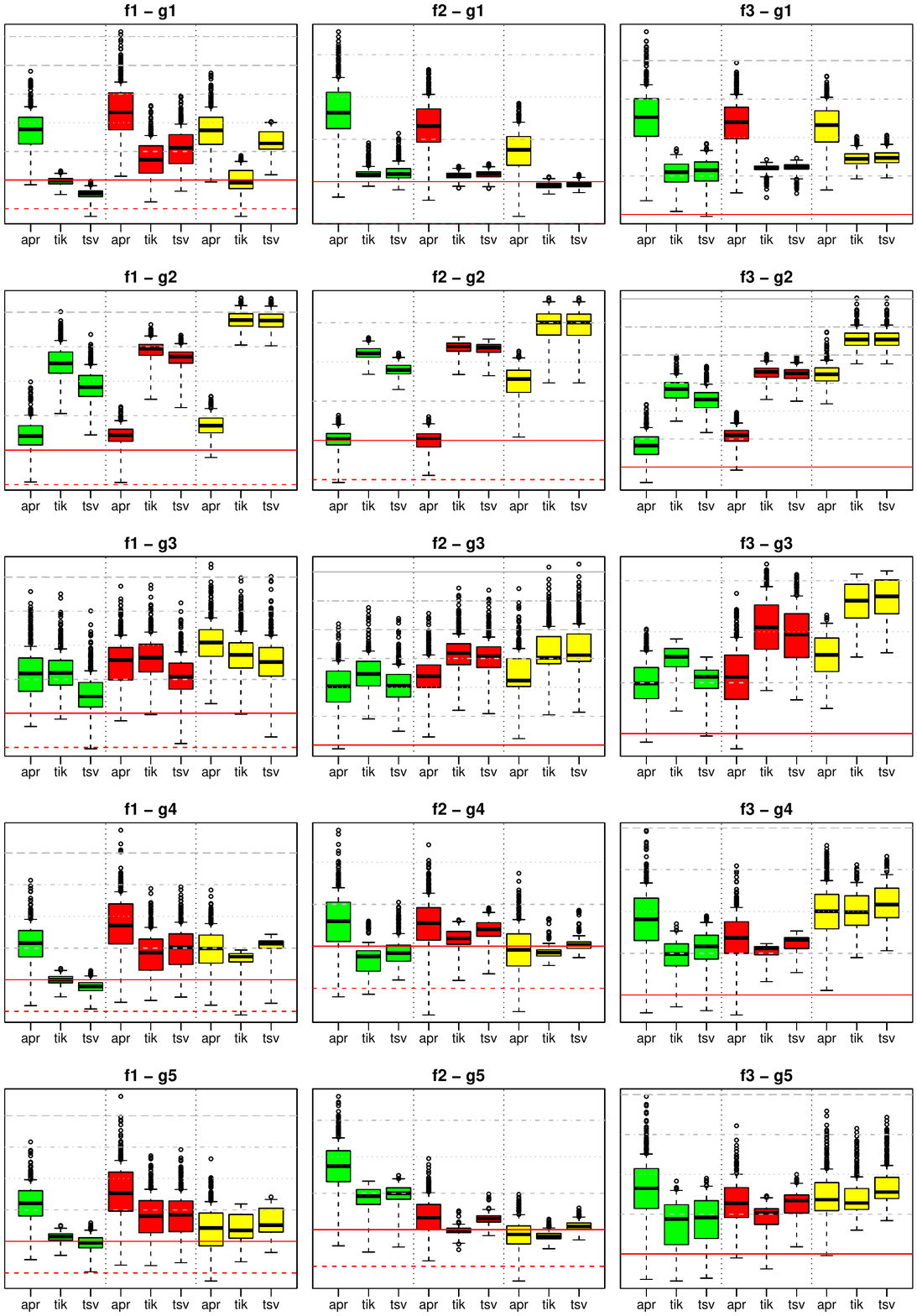}\vspace{-8mm}\]
\caption{\small \label{fig:lag risk ratio}  {\bf Box-plots of  $ISE(\widehat f)/ISE(\widehat f_{\mathrm{LAG}})$} 
for $\widehat f$ being $\widehat f_{\mathrm{APR}}$, $\widehat f_{\mathrm{TIKH}}$, $\widehat f_{\mathrm{tSVD}}$ 
with $n=100$ and noise levels $\sigma_0(g_j)/2^i$ for $i=1$ (green), $i=3$ (red) and $i=5$ (yellow). 
The box-plots are constructed on the basis of  400 simulation runs. In each column,   kernels $g$ vary from $g_1$ (top) to $g_5$ (bottom). 
In each row,   unknown function $f$ vary from $f_1$ (left) to $f_3$ (right). The empirical risk ratios are represented 
on a $\log_{10}$-scale: horizontal lines provide the references to the decibels (dB).
The plain red line, showing 0dB,  corresponds  to the equal error for our estimator and its competitor.
All values above this line suggest that our estimator has a smaller error. Other horizontal dashed 
lines provides positive (grey) or negative (red) increments of 1dB.
}
\end{figure}

\paragraph{Setting 1:  $g$ exactly known. } 
Figure \ref{fig:lag risk ratio} presents the box-plots of the ratios $ISE(\widehat f)/ISE(\widehat f_{\mathrm{LAG}})$,
constructed on the basis of    400 simulation runs, for $\widehat f$ being $\widehat f_{\mathrm{APR}}$, 
$\widehat f_{\mathrm{TIKH}}$ and $\widehat f_{\mathrm{tSVD}}$, $n=100$ and noise levels $\sigma_0(g_j)/2^i$ for $i=1,3,5$.  
 The empirical risk ratios are represented 
on a $\log_{10}$-scale: horizontal lines provide the references to the decibels (dB).
The plain red line, showing 0dB,  corresponds  to the equal error for our estimator and its competitor.
All values above this line suggest that our estimator has a smaller error.
The box-plots confirm that, except for a few rare cases (where the ratio is very close to one ), our estimator 
outperforms its competitors for all choices of kernels and test functions, and for all sample sizes and all noise levels.
Similar results were obtained for other noise levels and for $n=250$.

\begin{figure}
%\[ \includegraphics[width=16.5cm]{Fig-APR-LAG-n100-i1-P2-c2.pdf}\vspace{-12mm}\]
\[ \includegraphics[width=16.5cm]{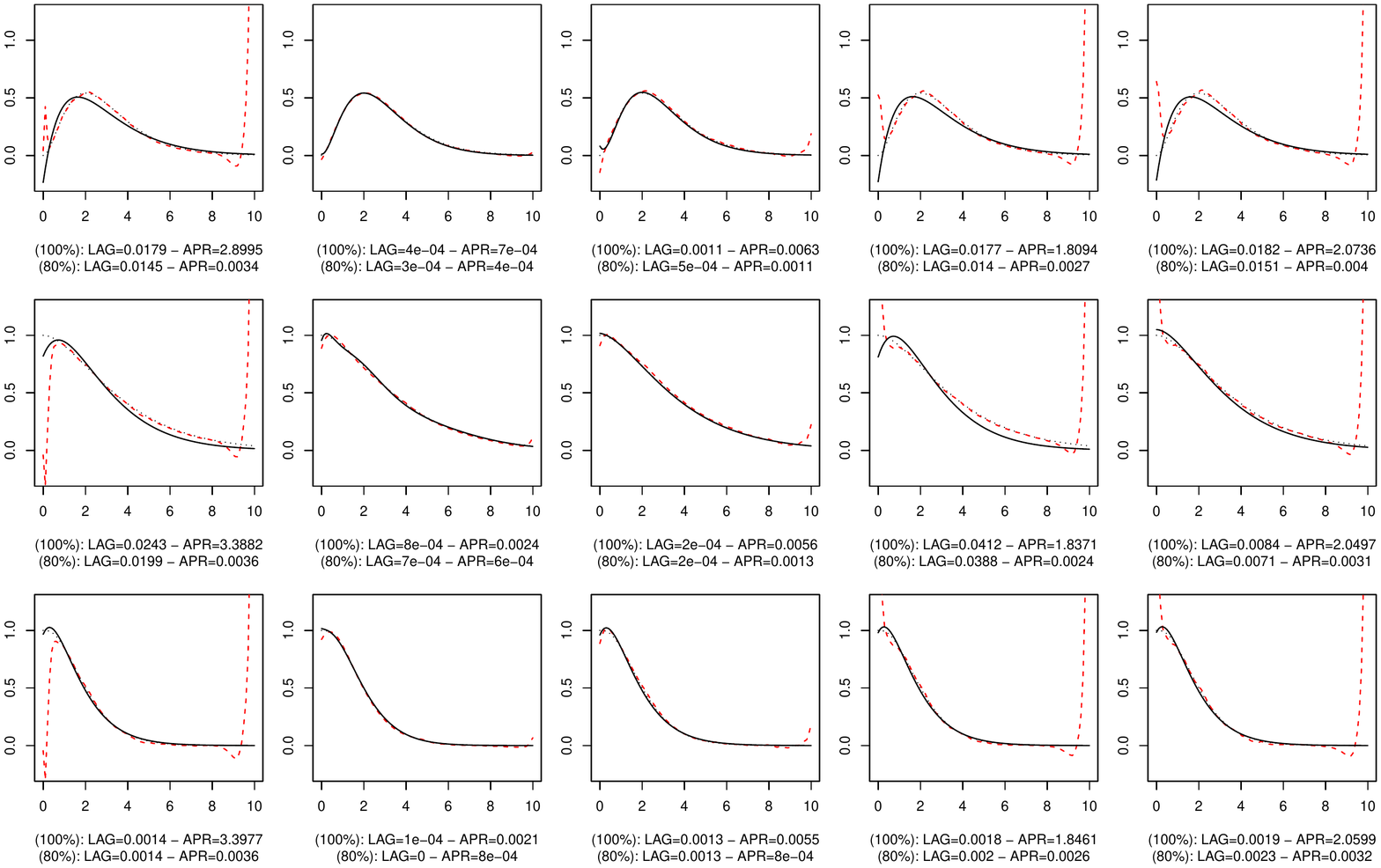}\vspace{-12mm}\]
\caption{\small \label{fig:lag example}  {\bf  Graphs of $\widehat f_{\mathrm{APR}}$ and   $\widehat f_{\mathrm{LAG}}$. }
One sample comparison with noise level $\sigma_0(g_j)/2$ and sample size $n=100$.
In each row,   kernels $g$ vary from $g_1$ (left) to $g_5$ (right). 
In each column,   unknown function $f$ vary from $f_1$ (top) to $f_3$ (bottom).
In each sub-figure, the dotted line is the unknown test function, the dashed line is $\widehat f_{\mathrm{APR}}$ 
and the plain line is $\widehat f_{\mathrm{LAG}}$. Below each sub-figure, we provide the values of 
$ISE(\widehat f_{\mathrm{LAG}})$ and $ISE(\widehat f_{\mathrm{APR}})$  
over the whole interval $[0,T]$ and also over 80\% of its interior points.} 
\end{figure}

Figure \ref{fig:lag example} presents the graphs of  $\widehat f_{\mathrm{LAG}}$ together  with $\widehat f_{\mathrm{APR}}$.
 for all test functions and kernels $g_j$ when the noise level is $\sigma_0(g_j)/2$.
It is easy to see that, in all cases, $\widehat f_{\mathrm{LAG}}$ shows a much more stable behavior than $\widehat f_{\mathrm{APR}}$ on the 
boundaries. The figures report the values of the ISE over the whole interval $[0,T]$ and also over 80\% of its interior points.
The overall error of $\widehat f_{\mathrm{LAG}}$ is always overwhelmingly smaller than that of $\widehat f_{\mathrm{APR}}$.
In the interior of the interval,  $\widehat f_{\mathrm{APR}}$  is competitive but the errors are extremely small for both estimators.
We also remind that $\widehat f_{\mathrm{APR}}$ cannot handle the case when $g$ is not known exactly and, hence, is not used for comparisons 
in Setting 2. 

To conclude this first set of simulations, Table \ref{tab:risks} in the Appendix
 provides the average values of   $ISE(\widehat f_{\mathrm{LAG}})$ computed over 400 simulation runs 
together with their standard deviations (in italic).

\begin{figure}
%\[ \includegraphics[width=16cm]{DCE-examples.pdf}\]
\[ \includegraphics[width=16cm]{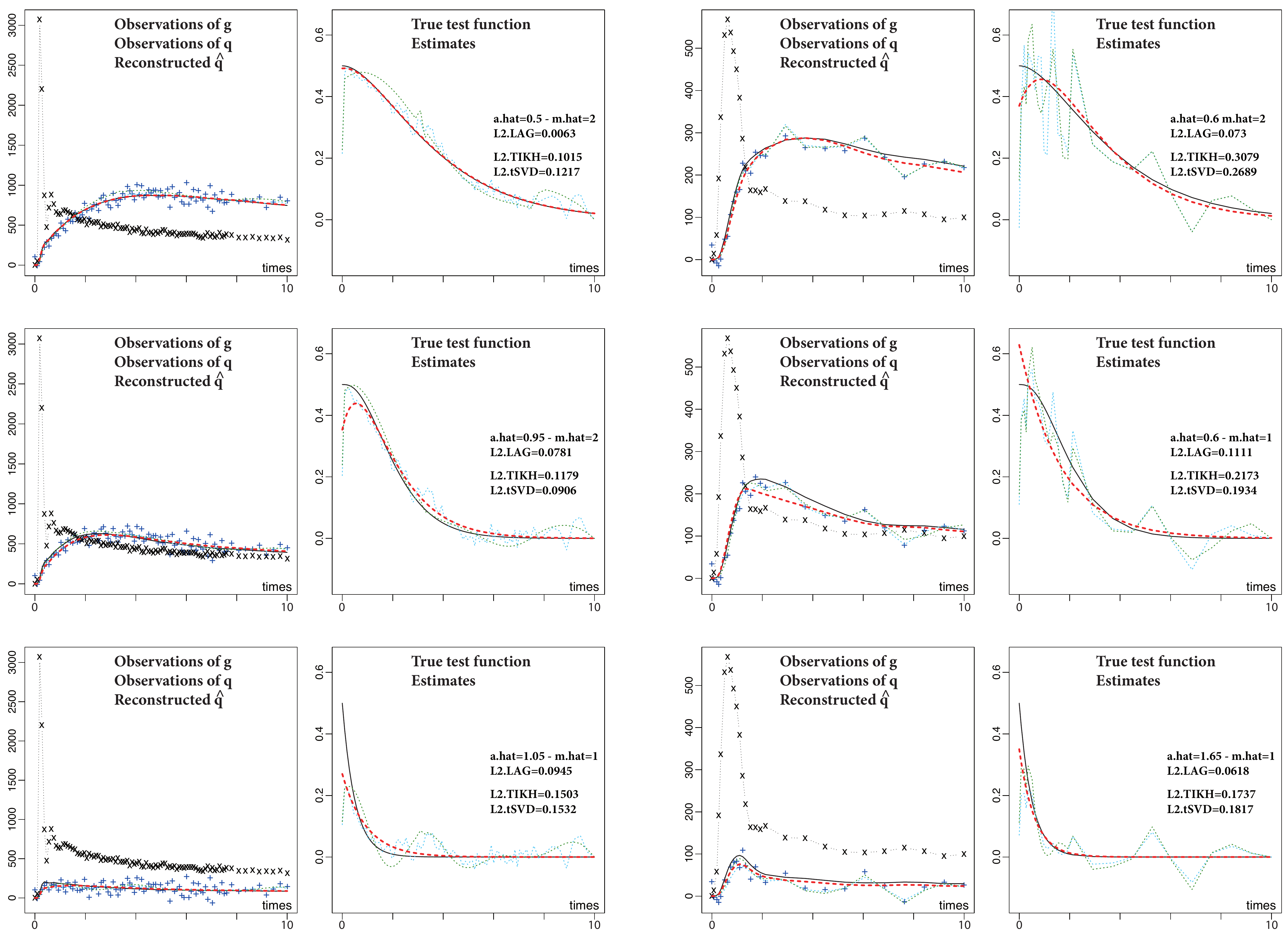}\]
\caption{\small \label{fig:dce example}  {\bf Example of estimations in the DCE setting experiment}: 
Left and right column corresponds to, respectively,  $g_{\mathrm{MRI}}$ and  $g_{\mathrm{CT}}$. 
From top to bottom, rows correspond to, respectively,  $f_2$, $f_3$ and $f_4$. In each column, the sub-figures on the left 
represent  observations $g(t_i)$ (black crosses $\times$) and $q(t_i)$ (blue crosses $+$) for $i=1,\ldots,n$, 
the  unknown $q$ (plain black line) and reconstructions $\widehat q_{\mathrm{LAG}}$ (dashed red line), 
$\widehat q_{\mathrm{TIKH}}$ (cyan dotted line) and $\widehat q_{\mathrm{tSVD}}$ (dotted green line).
The sub-figures on the right  represent  the true unknown test function $f$ (plain black line) and its estimators
$\widehat f_{\mathrm{LAG}}$ (dashed red line), $\widehat f_{\mathrm{TIKH}}$ (dotted cyan line) 
and  $\widehat f_{\mathrm{tSVD}}$ (plain green line).} 
\end{figure}

\paragraph{Setting 2: observations of  $g$ are available.} 
In this setting, we compare performances of $\widehat f_{\mathrm{LAG}}$, $\widehat f_{\mathrm{TIKH}}$ and  $\widehat f_{\mathrm{tSVD}}$.
 Figure \ref{fig:dce example} shows the graphs of the estimators obtained 
  for each combination of kernel $g$ and each test function $f$. 
Left and right columns correspond  to, respectively,  $g_{\mathrm{MRI}}$ and  $g_{\mathrm{CT}}$.
From top to bottom, rows correspond to  $f_2$, $f_3$ and $f_4$.  
In each column, the sub-figure   on the left shows the values of  $g(t_i)$ and $q(t_i)$ for $i=1,\ldots,n,$ 
together with the reconstructed estimator $\widehat q$ obtained by convolution of $\widehat f$ and $g$, while 
the sub-figure on the right, displays the test function $f$ together with $\widehat f_{\mathrm{LAG}}$, 
$\widehat f_{\mathrm{TIKH}}$ and $\widehat f_{\mathrm{tSVD}}$. Note that although the reconstructions  
$\widehat q$ based on $\widehat f_{\mathrm{LAG}}$, $\widehat f_{\mathrm{TIKH}}$ and $\widehat f_{\mathrm{tSVD}}$
are very similar, the precisions of estimators  $\widehat f_{\mathrm{LAG}}$, $\widehat f_{\mathrm{TIKH}}$ and $\widehat f_{\mathrm{tSVD}}$
themselves is dramatically different, especially in the case of $g_{\mathrm{MRI}}$ (the right two columns).
Note that $\widehat q$ is often used by radiologists as a visual indicator for the estimation quality. 
Figure~\ref{fig:dce example} demonstrates, however, that   this visual indicator is extremely poor and 
does not help in selection of an adequate   deconvolution procedure.

In addition, for the DCE imaging setting,  Figure \ref{fig:lag risk ratio} provides the box-plots of the ratios 
$ISE(\widehat f)/ISE(\widehat f_{\mathrm{LAG}})$, constructed over 400 simulation runs,  
for $\widehat f$ being $\widehat f_{\mathrm{TIKH}}$, $\widehat f_{\mathrm{tSVD}}$, 
together, with the average values of   $ISE(\widehat f_{\mathrm{LAG}})$ and their corresponding standard deviations. 
Figure \ref{fig:lag risk ratio}  confirms that again   $\widehat f_{\mathrm{LAG}}$ outperforms 
$\widehat f_{\mathrm{TIKH}}$ and  $\widehat f_{\mathrm{tSVD}}$ in all settings.

\begin{figure}
%\[ \includegraphics[width=12.2cm]{Real-Risks-DCE.pdf}\]
\[ \includegraphics[width=12.2cm]{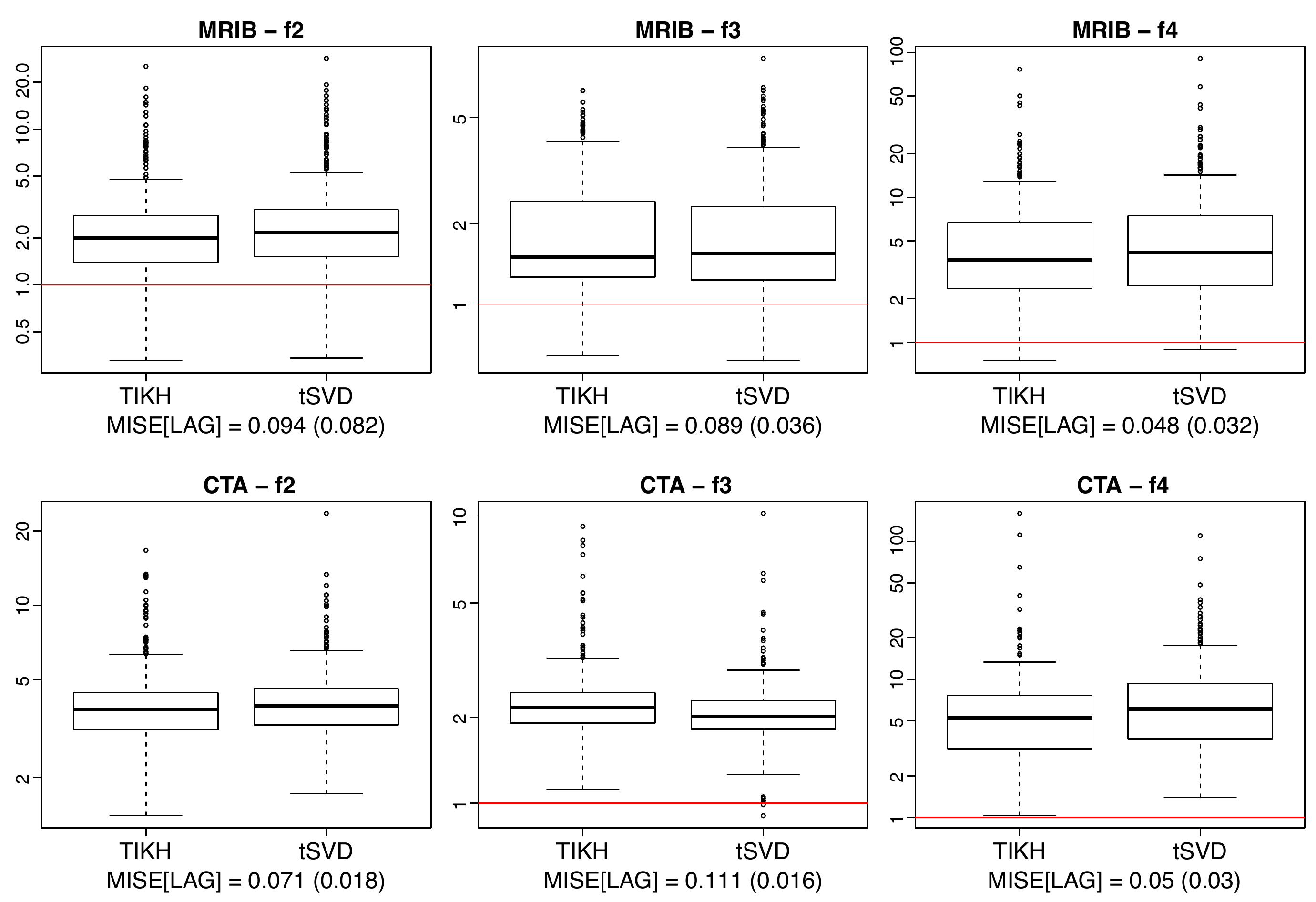}\]
\caption{\small \label{fig:ISE} {\bf Boxplot of the ratios $ISE(\widehat f)/ISE(\widehat f_{\mathrm{LAG}})$ 
for $\widehat f$ being $\widehat f_{\mathrm{TIKH}}$, $\widehat f_{\mathrm{tSVD}}$} constructed over 400 simulation runs. 
Upper row, $g_{\mathrm{MRI}}$ ($n=91$, $\sigma=75$). Lower row, $g_{\mathrm{CT}}$ ($n=28$, $\sigma=25$). 
From the left to the right column,  test functions $f_2$, $f_3$   and $f_4$. 
For each boxplot, the average value  of   $ISE(\widehat f_{\mathrm{LAG}})$ and its corresponding standard deviations
(in parenthesis)  are provided.}
\end{figure}

\begin{figure}
%\[ \includegraphics[width=12cm]{Real-Beta-DCE.pdf}\]
\[ \includegraphics[width=12cm]{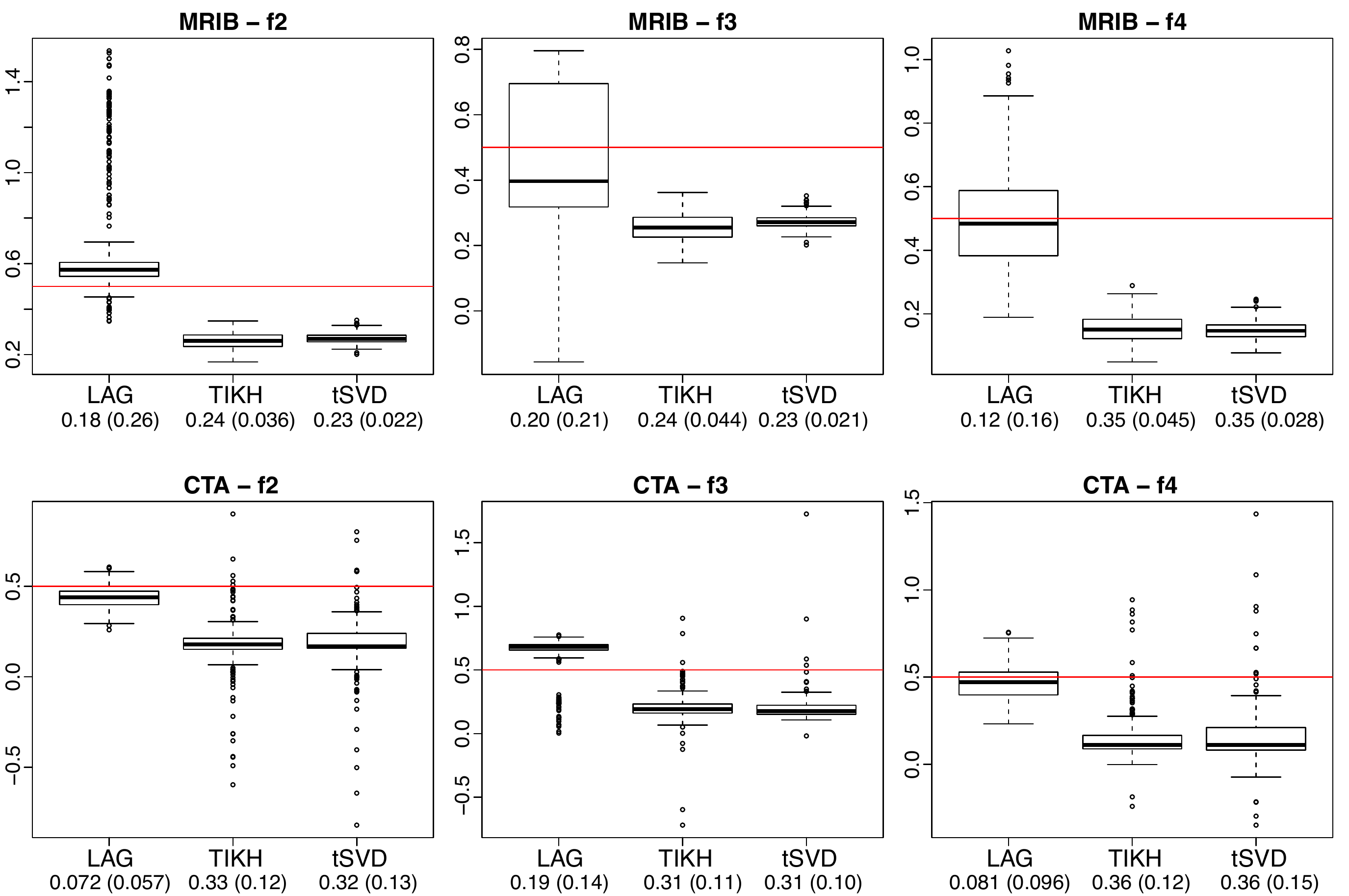}\]
\caption{\small \label{fig:beta}  {\bf Boxplot of $\widehat\beta=\widehat f(0)$} 
for $\widehat f$ being $\widehat f_{\mathrm{LAG}}$, $\widehat f_{\mathrm{TIKH}}$ and $\widehat f_{\mathrm{tSVD}}$
 computed over  400 simulation runs. 
Upper row: $g_{\mathrm{MRI}}$ ($n=91$, $\sigma=75$). Lower row: $g_{\mathrm{CT}}$ ($n=28$, $\sigma=25$). 
From the left to the right column: test functions $f_2$, $f_3$ and $f_4$. 
The red line corresponds to the true value of $\beta=0.5$ used in simulations. 
The average values of   $ISE(\widehat f_{\mathrm{LAG}})$ together with their corresponding standard deviations 
(in parentheses) are provided below each sub-figure.}
\end{figure}

Finally, Figure \ref{fig:beta} studies estimation of  $\beta$ in \fr{eq:AIF model}, the Tissue Blood Flow   
parameter which is of critical importance to radiologists and practitioners. Since  $1 - F(0)=1$, 
we use $\widehat\beta=\widehat f(0)$ as an estimator of $\beta$. Figure \ref{fig:beta} presents 
boxplots of the values of $\widehat \beta$ based on $\widehat f_{\mathrm{LAG}}$, $\widehat f_{\mathrm{TIKH}}$ and $\widehat f_{\mathrm{tSVD}}$ 
(constructed over 400 simulation runs).  The red line indicates the true   value $\beta=0.5$ used in simulations. 
Our estimator performs better than its competitors and shows  encouraging results for future applications in DCE imaging.

%%%%%%%%%%%%%%%%%%%%%%%%%%%%%%%%%%%%%%%%%%%%%%%%%%%%%%%%%%%%%%%%%%%%%%%%%%%%%%%%%%%%%%%%%%%%%%%%%%%

\section{Real-life experiments}
\label{sec:real_data}

In order to apply our procedure to real data, we used two  DCE-MRI sequences of 
one patient in  the REMISCAN cohort study \cite{remiscan}  who underwent 
anti-angiogenic therapy treatment for a metastatic renal carcinoma 
and showed positive response to the treatment after 3 months.
The first sequence has been obtained just before the start of the treatment and the second 15 days later.
One can notice that the first DCE-MRI sequence is more affected by the patient's movements: in spite of  being non-invasive, 
the first DCE-MRI experience is often stressful for a patient.

For each of $n = 91$ time instances, the DCE-MRI sequence is comprised  of  16 slices (or images) 
of $256 \times 256$ voxels that cover the metastases and surrounding areas. 
Injection of the contrast agent was administered so that the arrival of the contrast agent occurred after about 10 acquisition times. 
For each sequence, the measurements before the arrival of the contrast agent  were used to 
estimate the baseline image and its standard deviation $\sigma$. 
Then, the baseline was  removed from the sequence in order to obtain the enhancements. Extra times  before the arrival of the  contrast 
agent were removed from the series and time was shifted, so that $t_1=0$,  $AIF(t_1)\approx 0$ and $AIF(t_2)\gg 0$ 
and the effective sample size $n = 81$.  The time shift $\delta$ in \fr{eq:AIF model} was more or less constant   and 
was treated as  negligible for each sequence. Finally, we  set $AIF(t_1)=y(t_1)=0$ for $t<t_1$.
% before to start the estimation.

\begin{figure}
%\[ \includegraphics[width=15cm]{Enhancements-0108CP-14092009.pdf}\]
%\[ \includegraphics[width=15cm]{Enhancements-0108CP-29092009.pdf}\]
\[ \includegraphics[width=15cm]{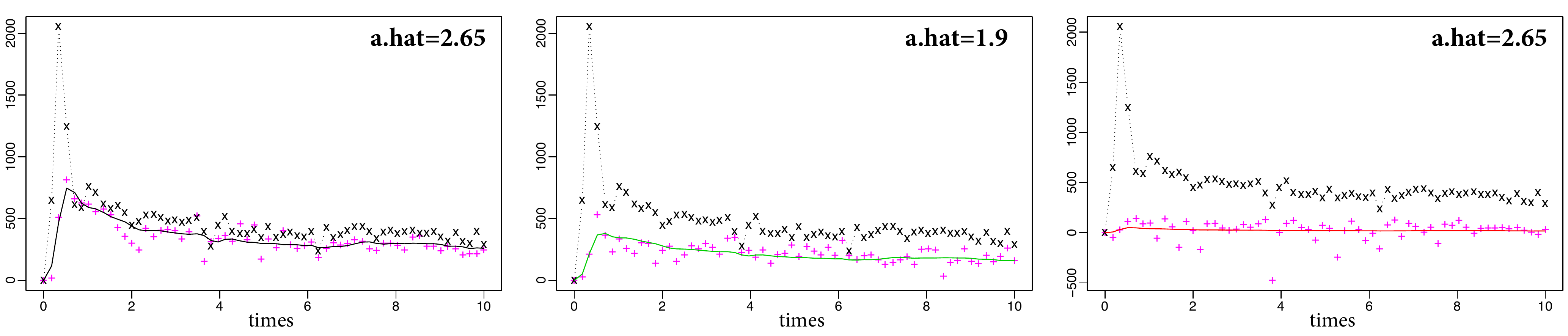}\]
\[ \includegraphics[width=15cm]{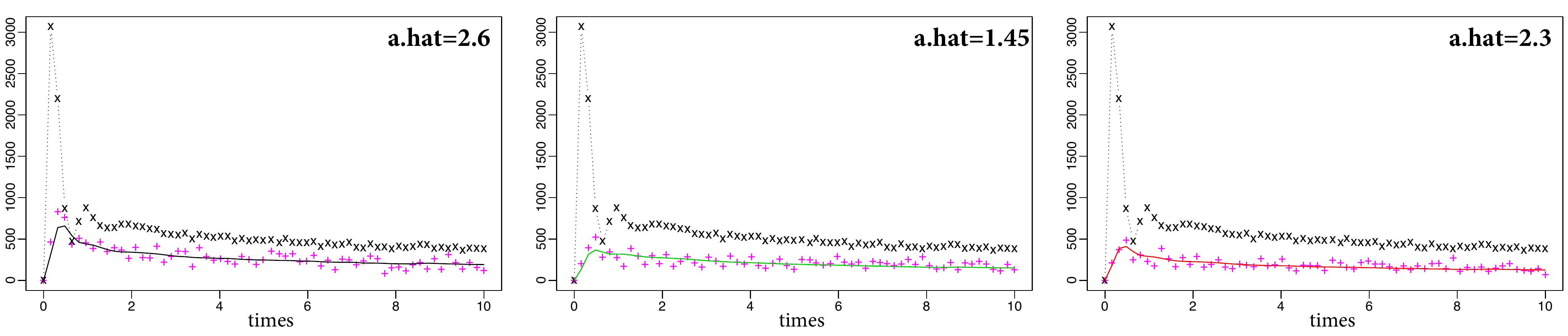}\]
\caption{\small \label{fig:enhancements}  {\bf The tissue enhancements (in the metastasis) (pink crosses $+$) 
and the corresponding AIFs (black crosses $\times$)} obtained from two DCE-MRI sequences of the same patient in the REMISCAN cohort study. 
% showing good response to the anti-angiogenic treatment after 3 months. 
% Enhancements were selected in the metastasis. 
Upper row corresponds to the DCE-MRI sequence obtained right before the start of the treatment;
bottom row  corresponds to the DCE-MRI sequence obtained 15 days after the treatment.
The value $\hat a$ chosen by cross-validation is provided in each sub-figure.  
The reconstructed enhancement  $\widehat q$ (plain line), obtained after estimation,
is displayed in each sub-figure as an indicator of the estimation precision.}
\end{figure}

In each sequence, we  selected three voxels inside the metastasis   and obtained three enhancement curves.
Since the aorta is visible on these DCE-MRI, its  images were  used for construction of estimators of  the $AIF$   
that were obtained as the average enhancements for all (around 400) voxels in the aorta. 
The six tissue enhancements as well as the two denoised $AIF$s are presented in the Figure \ref{fig:enhancements}.

\begin{figure}
\[ \includegraphics[width=11cm]{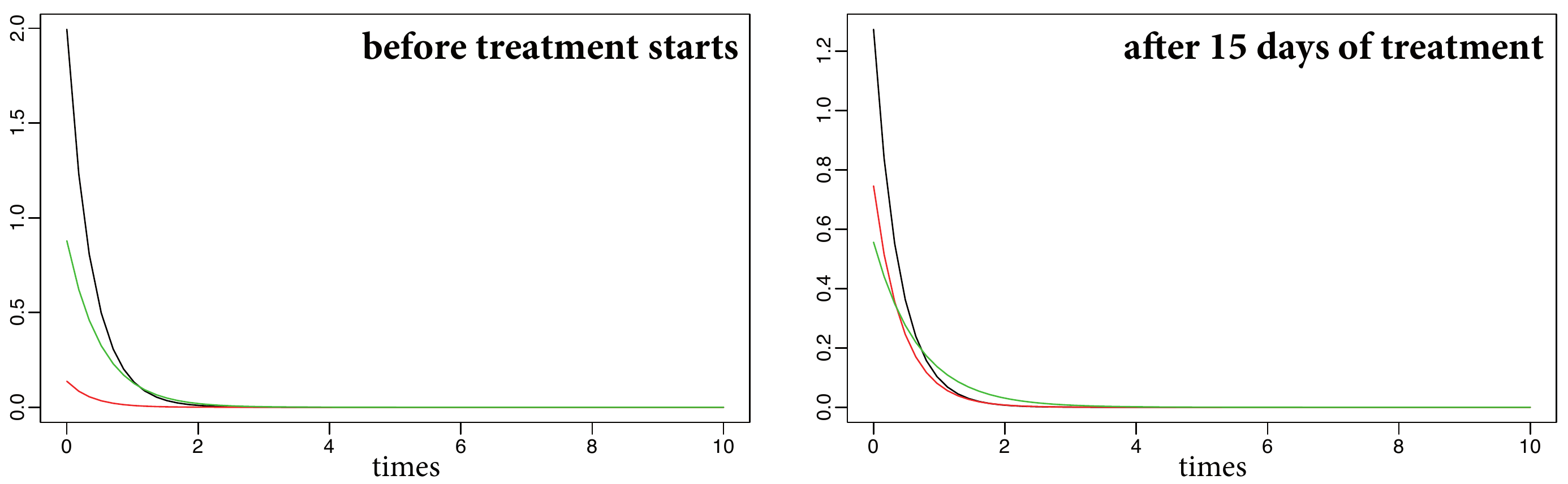}\]
%\[ \includegraphics[width=11cm]{Estimations-0108CP-14092009-29092009.pdf}\]
\caption{\small \label{fig:real-life DCE estimation}  {\bf DCE-MRI estimation results} 
The six estimates $\widehat f$ obtained from the  enhancements for six voxels selected in the two DCE-MRI sequences obtained  
before the start of the treatment starts (left) and after 15 days of the treatment (right). 
The colors of the estimates correspond  to the colors of the reconstructed enhancements in Figure \ref{fig:enhancements}. }
\end{figure}

The corresponding estimates are shown in Figure \ref{fig:real-life DCE estimation}. 
We remind that we estimate   function   $f=\beta(1-F)$ in \fr{eq:AIF model} 
where $(1-F)$ is  the survival function of the transit times of the contrast agent in the voxel 
and $\beta$ is the Tissue Blood Flow parameter  which can be estimated by $\widehat f(0)$.

%%%%%%%%%%%%%%%%%%%%%%%%%%%%%%%%%%%%%%%%%%%%%%%%%%%%%%%%%%%%%%%%%%%%%%%%%

Before the treatment  (upper line of Figure \ref{fig:enhancements}), the metastases exhibit  three 
different spatial  behaviors (hyper-vascular, vascular and necrotic) each illustrated by one of the three selected voxels: 
the enhancements correspond to the hyper-vascular (left),   vascular (center) and necrotic (right) area. 
After 15 days of treatment (bottom line of Figure \ref{fig:enhancements}), three new voxels have been selected, 
one  located in the hyper-vascular area observed before treatment (left), the two others   located in the vascular 
area observed before treatment (center and right) to check for reproducibility.

In the left panel of Figure \ref{fig:real-life DCE estimation}   corresponding to the   DCE-MRI sequence obtained before treatment, 
one can observe that the estimators  for  the hyper-vascular (black curve) and vascular (green curve) voxels show similar shapes 
and, hence, similar time transit distributions but strong differences in the estimated Tissue Blood Flow parameters. 
Moreover, the estimated Tissue Blood Flow parameter for the voxel   in the necrotic area (red curve) is, as expected, very small. 
In such poorly perfused tissues, one faces a small Signal to Noise Ratio (SNR) which challenges any deconvolution method.
However, as simulations show, our technique is relatively robust to low SNR values. 

In the right panel of Figure \ref{fig:real-life DCE estimation}  corresponding to the   DCE-MRI sequence obtained two weeks after the treatment, 
the estimator  for the voxel  located in the hyper-vascular area  (black curve in the right panel) 
shows very similar shape to the estimator  for the hyper-vascular voxel  observed before the treatment 
(black curve in the left panel) but a much lower estimated Tissue Blood Flow parameter.
This, however, can be expected as the result of the treatment which is aimed to reduce  the Tissue Blood Flow. 
The estimators obtained after the treatment for the two voxels located in the vascular area  look similar (red and green curves 
in the right panel) which is expected as they have been selected in a same area. Moreover, they also exhibit   clear reduction 
of the Tissue Blood Flow parameters compared to the estimator for the  vascular voxel before the treatment (green curve of left panel).

In conclusion, although we examined very limited experimental data, the estimators 
of the Tissue Blood Flow parameters and the survival functions show good reproducibility 
and are in accordance to what is expected by the clinicians.

%%%%%%%%%%%%%%%%%%%%%%%%%%%%%%%%%%%%%%%%%%%%%%%%%%%%%%%%%%%%%

%%%%%%%%%%%%%%%%%%%%%%%%%%%%%%%%%%%%%%%%%%%%%%%%%%%%%%%%%%%%%%%%%%%%%%%%%%%%%%%%%%%%%%%%%%%%%%%%%

\section{Discussion }
\label{sec:discussion}
\setcounter{equation}{0}

In the present paper, we study  a noisy version of the Laplace convolution equation.
Equations of this type frequently occur in various kinds of DCE imaging experiments.
We propose an estimation technique for the solutions of such equation based on the
expansion of the unknown solution, the kernel and the measured right-hand side over a system
of the Laguerre functions. The number of the terms in  the expansion  of the estimator
is controlled via complexity penalty. The technique  leads to an estimator with 
the risk within a logarithmic factor of the oracle risk.   

The major advantage of the methodology presented above is that it is practically usable, 
precise and computationally simple. Indeed, the expansion results in a small system of linear 
equations with the matrix of the system being triangular and Toeplitz. 
Therefore, the method is very easy and fast computationally  and produces no boundary effects 
due to the extension at zero and the cut-off at $T$.
In addition, application of the technique to discrete data does not require re-fitting the model
for each model size separately. On the contrary,  the vector of the Laguerre coefficients of the observed function
is fitted only once, for the largest model size, and then is truncated   for models of smaller sizes.
The complexity of representation of $g$ adjusts to the complexity of representation
of $f$ and the noise level. Moreover, if  $g$ can be represented by a finite expansion
over Laguerre functions with $k$ terms, the matrix of the system is $k$-diagonal.
The method performs very well in simulations. It is much more precise than the estimators described in  
 Abramovich {\it et al.}  (2013), as well as the estimators obtained by 
  the SVD and the Tikhonov regularization.

Another important property of the method   is that it can be easily applied when 
the kernel is not known exactly and is only observed at some points. 
This distinguishes the present technique from the approach of Abramovich {\it et al.}  (2013)
 which strongly depends on the precise knowledge of the analytic form of the kernel 
and, hence, cannot be applied to solution of real-life problems. 
In the paper, we describe application of our methodology to analysis of the DCE-MRI sequences
where the kernels are estimated on the basis of imaging data.

There are few more advantages which are associated with the use of Laguerre functions basis.
Since one important goal of future analysis of DCE data  is classification of the tissues
and clustering of curves $f(t) = \beta(1-F(t))$ which characterize their blood flow properties,
representation of the curves via Laguerre basis allows to replace the problem of classification
of curves by classification of relatively low-dimensional vectors.
In addition, due to the absence of boundary effects, the method allows to estimate  $\beta$,  
the Tissue Blood Flow parameter, which characterizes the micro-vascular properties of the tissue
and is of extreme interest to medical doctors.
Indeed, our simulations demonstrate that those estimators are fairly accurate.
However, since they are based on a global estimator $\widehat  f_{\mathrm{LAG}}$ 
rather than on a local estimator at zero, there is some room for improvement.
However, this issue is a matter of future investigation.

%%%%%%%%%%%%%%%%%%%%%%%%%%%%%%%%%%%%%%%%%%%%%%%%%%%%%%%%%%%%%%%%%%%%%%%%%%%%%%%%%%%%%%%%%%%%%%%%%%%%%%%%%%%%%%%%%%%%

\section*{Acknowledgments}

Marianna Pensky was partially supported by National Science Foundation
(NSF), grants  DMS-1106564 and DMS-1407475. The authors want to express sincere gratitude to Sergei Grudski
for his invaluable help in the proof of Lemma \ref{lem:rho_v}  and very helpful discussions.

%%%%%%%%%%%%%%%%%%%%%%%%%%%%%%%%%%%%%%%%%%%%%%%%%%%%%%%%%%%%%%%%%%%%%%%%%%%%%%%%%%%%%%%%%%%%%%%%%%%%%%%%%%%%5

%%%%%%%%%%%%%%%%%%%%%%%%%%%%%%%%%%%%%%%%%%%%%%%%%%%%%%%%%%%%%%%%%%%%%%%%%%%%%%%%%%%%%%%%%%%%%%%%%%%%%%%%%%%%5
%%^^^^^^^^^^^^^^^^^^^^^^^^^^^^^^^^^^^^^^^^^^^^^^^^^^^^^^^^^^^^^^^^^^^^^^^^^^^^^^^^^^^^^^^^^^^^^^^^^^^^^^^^^^^^^^^^^^

\section{Proofs }
\label{sec:proofs}
\setcounter{equation}{0}

\subsection{Proof of Proposition \ref{MISE}}

Recall that 
\begin{eqnarray} \label{eq:error_terms}
{\mathbb E}( \|\hat f_m-f\|^2_2)&=&
\|f-f_m\|_2^2+  {\mathbb E}(\|\widehat{{\mathbf f_m}}-{\mathbb E}(\widehat{{\mathbf f_m}})\|^2)
+  \|{\mathbb E}(\widehat{{\mathbf f_m}})-{\mathbf f_m}\|^2  
\end{eqnarray}
The second term in \fr{eq:error_terms} is the variance term which is equal to
\[
{\mathbb E}(\|\widehat{{\mathbf f_m}}-{\mathbb E}(\widehat{{\mathbf f_m}})\|^2)=
\sigma^2 T n^{-1} {\mathbb E} \left[(\mathbf A_m\vec \varepsilon \;)^T\mathbf A_m\vec \varepsilon\;\right]=
\sigma^2 T n^{-1} {\rm Tr}(\mathbf A_m^T\mathbf A_m)
\]
with $A_m$ defined in \fr{def:Am}.
% \begin{equation}\label{def:Am}
% \mathbf A_m = \sqrt{\frac nT}\, \mathbf G_m^{-1}\bJ_{m,M} (\Phi_M^T\;\Phi_M)^{-1}\Phi_M^T.
% \end{equation}
Observing that 
\begin{eqnarray*}
%\label{def:Qm}
  \bJ_{m,M}(T n^{-1} \Phi_M^T\Phi_M)^{-1}\bJ_{m,M}^T \mathbf  G_m^{-T}\mathbf G_m^{-1} &=&  
 \frac n T [(\Phi_M^T\Phi_M)^{-1}]_m (\bG_m\bG_m^T)^{-1}\\
 &=& \frac n T [(\Phi_M^T\Phi_M)^{-1}]_m ([\bG_M\bG_M^T]_m)^{-1} = {\mathbf Q_m}, \nonumber
\end{eqnarray*}
we obtain
\begin{equation}\label{eq:mainvar}
{\mathbb E}(\|\widehat{{\mathbf f_m}}-{\mathbb E}(\widehat{{\mathbf f_m}})\|^2)=
\sigma^2T n^{-1} {\rm Tr}({\mathbf Q_m}).
\end{equation}
% Note that $\mathbf Q_m = nT^{-1} [(\Phi_M^T\Phi_M)^{-1}]_m (\bG_m\bG_m^T)^{-1} = nT^{-1} [(\Phi_M^T\Phi_M)^{-1}]_m ([\bG_M\bG_M^T]_m)^{-1}$.\\
%$\mathbf Q_m= nT^{-1} [(\Phi_M^T\Phi_M)^{-1}]_m \mathbf G_m^{-T}\mathbf G_m^{-1} = \bOm_m (\bG_m\bG_m^T)^{-1}$
%
%
For the last term of right hand side of (\ref{eq:dec1}), we derive
$$
 \|{\mathbb E}(\widehat{{\mathbf  f_m}})-{\mathbf f_m}\|^2 =
\|\mathbf G_m^{-1}\bJ_{m,M} (\Phi_M^T\;\Phi_M)^{-1}\Phi_M^T(\vec q-\vec q_M)\|^2.
$$
Using inequality $\rho^2(\mathbf A)\leq {\rm Tr}(\mathbf A^T\mathbf A)=\|\mathbf A\|_2^2$,
and the fact that, for all $k\in {\mathbb N}$ one has $\|\phi_k\|_\infty \leq \sqrt{2a}$
(see Abramowitz and Stegun  (1964), 22.14.12 and following), we bound this term by
\begin{eqnarray}
 \|{\mathbb E}(\widehat{{\mathbf  f_m}})-{\mathbf f_m}\|^2 &\leq & \rho^2(\mathbf G_m^{-1}\bJ_{m,M}
(\Phi_M^T\;\Phi_M)^{-1}\Phi_M^T) \|\vec q-\vec q_M\|_{n}^2\nonumber\\
&\leq & \frac Tn {\rm Tr}({\mathbf Q_m}) \sum_{i=1}^n (q(t_i)-q_M(t_i))^2\label{the_condition}\\
&\leq & \frac Tn {\rm Tr}({\mathbf Q_m}) \sum_{i=1}^n \left(\sum_{k=M+1}^{\infty}q^{(k)}\phi_k(t_i)\right)^2\nonumber\\
&\leq & \frac Tn {\rm Tr}({\mathbf Q_m})(2a n)\left(\sum_{k=M+1}^{\infty}|q^{(k)}|\right)^2.\nonumber
\end{eqnarray}
Using Assumption  {\bf (A0)},  we obtain that 
$$
\left(\sum_{k=M+1}^{\infty}|q^{(k)}|\right)^2= \left(\sum_{k=M+1}^{\infty}k^2|q^{(k)}|
k^{-2}\right)^2 \leq  C_q \sum_{k=M+1}^{\infty} k^{-4}\leq C_q M^{-3}/3,
$$
so that
\begin{equation}\label{eq:boundres}
\|{\mathbb E}(\widehat{{\mathbf f_m}})-{\mathbf f_m}\|^2
\leq \frac{ 2aT C_q}{3} \frac{{\rm Tr}({\mathbf Q_m})}{M^{3}}.
\end{equation}
Combination of  (\ref{eq:error_terms}), (\ref{eq:mainvar}) and (\ref{eq:boundres}) completes the proof
of \fr{eq:risk}, provided that $M= n^{(1+\eta)/3}$ and $n\geq (2aC_q/\sigma^2)^{1/\eta}$.

%%%%%%%%%%%%%%%%%%%%%%%%%%%%%%%%%%%%%%%%%%%%%%%%%%%%%%%%%%%%%%%%%%%%%%%%%%%%%%%%%%%%%%%%%%%%%%%%%%%%%

\subsection{Proof of Theorems \ref{th:risk_bound}  and \ref{th:asymp_risk} }

\noindent{\bf Proof  of Theorem  \ref{th:risk_bound}. }
For $m \in \mathM_n$, we consider the associated subspaces
$S_m \subseteq R^M$ defined by
$$
\bt \in S_m\ \ \mbox{if}\ \   \bt = (t^{(0)}, t^{(1)}, \ldots, t^{(m-1)}, 0,0, \ldots, 0)^T.
$$
For ${\mathbf t}\in S_m$, a vector of ${\mathbb R}^{M}$ with at most its first $m$ coordinates non-zero, $\widehat{{\mathbf f_m}}$ can be related to the minimizer over $S_m$ of the contrast
$$\gamma_n({\mathbf t})= \|{\mathbf t}\|^2 -2 \langle {\mathbf t},  \mathbf G_M^{-1}\widehat{{\mathbf q_M}}\rangle.$$
Note that, for ${\mathbf t}\in S_m$,
$$ \langle {\mathbf t},  \mathbf G_M^{-1}\widehat{{\mathbf q_M}}\rangle=  \langle [{\mathbf t}]_m,  \mathbf G_m^{-1}[\widehat{{\mathbf q_M}}]_m\rangle$$
where we recall that $[x]_m$ is the $m\times 1$ vector obtained by retaining the $m$ first coordinates of $x$.
Reciprocally, let us denote by  ${\mathbf q}_{m,M}$,  ${\mathbf f}_{m,M}$,  and $\widehat{{\mathbf f}}_{m,M} $ the $M$-dimensional vectors
where the first $m$ elements coincide with the elements of $m$-dimensional vectors $\bq_m$, $\bof_m$,   and $\widehat{\bof}_m$ respectively, and the last $(M-m)$ elements are identical zeros. Since, for ${\mathbf t}\in S_m$, we have
$\gamma_n({\mathbf t})=\|{\mathbf t}- \mathbf G_M^{-1}\widehat{{\mathbf q_M}}\|^2-\| \mathbf G_M^{-1}\widehat{{\mathbf q_M}}\|^2$, we can see that clearly $$\widehat{\bof}_{m,M} =\arg\min_{{\mathbf t}\in S_m} \gamma_n({\mathbf t}) \mbox{ and } \widehat{{\mathbf f}_{m}}=[\widehat{{\mathbf f}}_{m,M}]_m.$$
Now since $\gamma_n(\widehat{\bof}_{m,M} )=-\|\widehat{\bof}_{m,M}\|^2=-\|\widehat{\bof}_m\|^2$, we can see that
 \begin{equation}\label{hatmgam}
 \widehat{m}=\arg \min_{m\in {\mathcal M}_n}
\left\{ \gamma_n(\widehat{\bof}_{m,M}) + {\rm pen}(m)\right\},
 \end{equation}

Let $m, m' \in \mathM_n$, $\bt \in S_{m'}$ and  $\bs \in S_m$ and observe that
\begin{eqnarray}\label{obs1}
\gamma_n(\bt)-\gamma_n(\bs) &=&
\| \bt-{\mathbf f_M}\|^2 -\|\bs-{\mathbf f_M}\|^2 -2 \langle \bt-\bs, \bG_{M}^{-1} \widehat{{\mathbf q_M}}-{\mathbf f_M} \rangle,
\end{eqnarray}
where ${\mathbf f_M}$ is the vector of the true $M$ first Laguerre coefficients of function $f$.
Note that, due to orthonormality of the Laguerre system, for any $m$,
\begin{equation}\label{obs2}
\|\hat{f}_m  - f\|^2_2 = \| \widehat{\bof}_{m,M} - {\mathbf f_M} \|^2 + \sum_{j=M}^\infty \lkr f^{(j)}\rkr^2
\;  \mbox{ and } \;
\|f_m  - f\|^2_2 = \| \bof_{m,M} - {\mathbf f_M} \|^2 + \sum_{j=M}^\infty \lkr f^{(j)}\rkr^2.
\end{equation}

% Since $M$ is large, the sum in the relation above is infinitesimally small in comparison with
% $\| \widehat{\vbof}_m - \vf \|^2$, so that in what follows, we shall treat  $\|\hat{f}_m  - f\|^2$
% and $\| \widehat{\vbof}_m - \vf \|^2$ as equal interchangeable quantities.

Now, the definition of $\widehat m$ as given by (\ref{hatmgam}) yields that for any $m \in {\mathcal M}_n$ one has
$$
\gamma_n(\widehat{\bof}_{\widehat{m}, M}) + \pen (\widehat m) \leq \gamma_n(\bof_{m,M})+  \pen (m),
$$
which with (\ref{obs1}), implies
\begin{eqnarray*} \| \widehat{\bof}_{\widehat{m},M} - {\mathbf f_M} \|^2  &\leq &   \| \bof_{m,M} - {\mathbf f_M} \|^2+ \pen(m)  + 2 \langle \widehat{\bof}_{\widehat{m},M} - \bof_{m,M}, \bG_{M}^{-1} \widehat{{\mathbf q_M}}-{\mathbf f_M} \rangle -\pen(\widehat{m})\\ &\leq &  \| \bof_{m,M} - {\mathbf f_M} \|^2+ \pen(m) + 2\|\widehat{\bof}_{\widehat{m},M} - \bof_{m,M}\| \sup_{\bt \in S_{\hat m\vee m} } | \langle \bt, \bG_{M}\widehat{{\mathbf q_M}}-{\mathbf f_M} \rangle |- \pen(\widehat{m}).
\end{eqnarray*}
% Therefore, using (\ref{obs2}), we obtain that, for any $m \in {\mathcal M}_n $
% \be \label{eq:risk1}
% \|\hat{f}_{\widehat{m}} - f\|_{2}^2 \leq  \|f_m - f\|_{2}^2 + \pen(m) + 2\|\widehat{\bof}_{\widehat{m},M} - \bof_{m,M}\| \sup_{\stackrel{\bt \in S_{\hat m\vee m}}{\|\bt \|=1} } | \langle \bt, \bG_{M}^{-1} \widehat{{\mathbf q_M}}-{\mathbf f_M} \rangle | - \pen(\widehat{m}).
% \ee
Due to $2xy\leq (x^2/\theta)+ \theta y^2$ for all $x>0, y>0$ and $\theta>0$, we get, choosing 
$\theta=2$
\begin{eqnarray}\nonumber\| \widehat{\bof}_{\widehat{m},M} - {\mathbf f_M} \|^2& \leq & \| \bof_{m,M} - {\mathbf f_M} \|^2 + \pen(m) + \frac 12\|\widehat{\bof}_{\widehat{m},M} - \bof_{m,M}\|^2 \\  \label{s0}  && + 2 \sup_{\stackrel{\bt \in S_{\hat m\vee m}}{\|\bt \|=1} }  \langle \bt, \bG_{M}^{-1} \widehat{{\mathbf q_M}}-{\mathbf f_M} \rangle^2 - \pen(\widehat{m}).
\end{eqnarray} 
Due to $|x+y|^2\leq (1+\theta)x^2 + (1+\theta^{-1})y^2$ for all $x,y$ and $\theta>0$, we get 
choosing $\theta=3$,
\begin{equation}\label{s1}
\|\widehat{\bof}_{\widehat{m},M} - \bof_{m,M}\|^2\leq \frac 43 \|\widehat{\bof}_{\widehat{m},M} -  {\mathbf f_M}\|^2 + 4 \|\bof_{m,M}-f_M\|^2 ,\end{equation}
and, choosing $\theta=2$,
\begin{equation}\label{s2} 2 \sup_{\stackrel{\bt \in S_{\hat m\vee m}}{\|\bt \|=1} }  \langle \bt, \bG_{M}^{-1} \widehat{{\mathbf q_M}}-{\mathbf f_M} \rangle^2 - \pen(\widehat{m})\leq  \Delta_{m,\widehat m}^{(1)} +  \Delta_{m,\widehat m}^{(2)}
\end{equation}
where
$$\Delta_{m, \widehat m}^{(1)} =  3\sigma^2 \sup_{\stackrel{\bt \in S_{\hat m\vee m}}{\|\bt \|=1}} \langle \bt , \bG_M^{-1} (\Phi_M^T\Phi_M)^{-1}\Phi_M^T\vec\varepsilon \;\rangle^2 - \frac 34 \pen(\widehat{m}),$$
$$\Delta_{m, \widehat{m} }^{(2)} = 6\sup_{\stackrel{\bt \in S_{\hat m\vee m}}{\|\bt \|=1} }\langle \bt , \bG_M^{-1}(\Phi_M^T\Phi_M)^{-1}\Phi_M^T(\vec q-\vec q_M)\rangle^2 - \frac 14 \pen(\widehat{m}).$$
Plugging (\ref{s1}) and (\ref{s2}) into (\ref{s0}) yields
$$
\frac 13\|\widehat{\bof}_{\widehat{m},M} -  {\mathbf f_M}\|^2  \leq  3\|\bof_{m,M}- {\mathbf f}_M\|^2  + \pen(m) + \Delta_{m,\widehat m}^{(1)} +  \Delta_{m,\widehat m}^{(2)},
$$
Using (\ref{obs2}), we obtain 
\begin{equation}\label{s3} \frac 13 \|\hat{f}_{\widehat{m}} - f\|_{2}^2 \leq 3 \|f_m - f\|_{2}^2 + \pen(m) + \Delta_{m,\widehat m}^{(1)} +  \Delta_{m,\widehat m}^{(2)},\end{equation}

\noindent Now we  have 
\begin{eqnarray*}
\Delta_{m, \widehat m}^{(2)} & =  &  6 \sup_{\stackrel{\bt \in S_{\hat m\vee m}}{\|\bt \|=1} }\langle \bt ,  \bG_{m\vee \hat m}^{-1}J_{m\vee \hat m, M}(\Phi_M^T\Phi_M)^{-1}\Phi_M^T(\vec q-\vec q_M)\rangle^2 - \frac 14\pen(\widehat{m})\\ &\leq & 6\|\bG_{m\vee \hat m}^{-1}J_{m\vee \hat m, M}(\Phi_M^T\Phi_M)^{-1}\Phi_M^T(\vec q-\vec q_M)\|^2-\frac 14\pen(\widehat{m})\end{eqnarray*}
and  under {\bf (A0)}, $M=n^{(1+\eta)/3}$ and $n\geq (2aC_q/\sigma^2)^{1/\eta}$, we get, as in (\ref{eq:boundres}), 
\begin{eqnarray*}6\|\bG_{m\vee \hat m}^{-1}J_{m\vee \hat m, M}(\Phi_M^T\Phi_M)^{-1}\Phi_M^T(\vec q-\vec q_M)\|^2 &\leq &2\frac Tn \sigma^2 {\rm Tr}({\mathbf Q_{m\vee \widehat m}})\leq 2\frac Tn \sigma^2 ({\rm Tr}({\mathbf Q_m})+ {\rm Tr}({\mathbf Q_{\widehat{m}}})) \\ &\leq & \frac 14(\pen(m)+\pen(\widehat{m})).
\end{eqnarray*}
As a consequence \begin{equation}\label{delta2} \Delta_{m, \widehat m}^{(2)}\leq \frac 14 \pen(m).\end{equation}

\noindent Now, denote
\be \label{eq:tau}
\tau(m, m') = 2\frac{T}{n} \lkv  v^2_{m^*} + \log( ( m^*\rho_{m^*}/\rho_1)^{2\kappa} ) \rho^2_{m^*} \rkv,
\ee
where $m^* = m \vee m'$.  Then
\be \label{eq:Deltam}
 \Delta_{m,\widehat m}^{(1)}  \leq   3  \sigma^2 \lkv \sup_{\bt \in S_{m \vee \whm}}  \langle \bt, \bG_M^{-1} (\Phi_M^T\Phi_M)^{-1}\Phi_M^T\vec\varepsilon \; \rangle^2 - \tau(m,\whm) \rkv_+ %\nonumber\\
 + 3 \sigma^2\tau(m,\whm) - \frac 34\pen(\whm).
\ee
Using the fact that  $3 \sigma^2 \tau(m,\whm) \leq (3/4)(\pen(m)+ \pen(\whm))$,  combining
\fr{s3},  \fr{delta2} and \fr{eq:Deltam}, we derive
$$\frac 13 \|\hat{f}_{\widehat{m}} - f\|_{2}^2 \leq 3 \|f_m - f\|_{2}^2 + 2\pen(m) + 3 \sigma^2 \lkv \sup_{\bt \in S_{m \vee \whm}}  \langle \bt, \bG_M^{-1} (\Phi_M^T\Phi_M)^{-1}\Phi_M^T\vec\varepsilon \; \rangle^2 - \tau(m,\whm) \rkv_+ ,
$$
We obtain
\beqn \label{eq:MISE}
\|\hat{f}_{\widehat{m}} - f\|^2_2 & \leq & 9\;  \|f_m - f\|^2_2  + 6 \; \pen(m) \nonumber \\
& + &
9 \sigma^2 \lkv \sup_{\bt \in S_{m \vee \whm}}  \langle \bt, \bG_M^{-1}(\Phi_M^T\Phi_M)^{-1}\Phi_M^T\vec\varepsilon \;
\rangle^2 - \tau(m,\whm) \rkv_+ .
\eeqn
Hence, validity of Theorem \ref{th:risk_bound} rests on the  following lemma which will be proved later.

\begin{lemma} \label{lem:large_deviation}
Under the assumptions of Theorem \ref{th:risk_bound}, for any $m\geq 1$, one has
$$\EE \lkv \sup_{\bt \in S_{m \vee \whm}, \|\bt\|=1}  \langle \bt, \bG_M^{-1} (\Phi_M^T\Phi_M)^{-1}\Phi_M^T\vec\varepsilon \; \rangle^2 - \tau(m,\whm) \rkv_+  \leq
\frac{8 T \rho_1^2}{mn}.
$$
\end{lemma}
Proof of  Lemma  \ref{lem:large_deviation} is given in Appendix \ref{sec:supplementary}.
\\

\noindent{\bf Proof  of Theorem  \ref{th:asymp_risk}. }
 Let $m_0 = \arg\min_m [\|f_m - f\|_{2}^2 + \sigma^2 T n^{-1}\, \log(m) v_m^2]$.
Then,  due to bounds \fr{eq:vm_bounds} on  $v_m^2$, one has $ m_0 \rightarrow \infty$ and
$ (m_0^{2r+1} T)/n \rightarrow 0$ as $T/n \rightarrow 0$. Hence,
it also follows from \fr{eq:vm_bounds}(see Lemma \ref{lem:rho_v}) that  $\rho_m^2  \log(m^2\rho_m^2) \propto \log(m) \, v_m^2$   as
$m \rightarrow \infty$ which, in combination with Theorem \ref{th:risk_bound},
completes the proof.
\\

%%%%%%%%%%%%%%%%%%%%%%%%%%%%%%%%%%%%%%%%%%%%%%%%%%%%%%%%%%%%%%%%%%%%%%%%%%%%%%%%%%%%%%%%%%%%%%%%%%%%%%%%%%%%5
%%^^^^^^^^^^^^^^^^^^^^^^^^^^^^^^^^^^^^^^^^^^^^^^^^^^^^^^^^^^^^^^^^^^^^^^^^^^^^^^^^^^^^^^^^^^^^^^^^^^^^^^^^^^^^^^^^^^

%%%%%%%%%%%%%%%%%%%%%%%%%%%%%%%%%%%%%%%%%%%%%%%%%%%%%%%%%%%%%%%%%%%%%%%%%%%%%%%%%%%%%%%%%%%%%%%%%%%%%%%%%%%%

\noindent
Fabienne Comte \\
Sorbonne Paris Cit\'{e}\\
Universit\'{e}  Paris  Descartes,\\
 MAP5, UMR CNRS 8145, France\\
{\em fabienne.comte@parisdescartes.fr}\\

\noindent
Charles-Andr\'{e} Cuenod\\
Sorbonne Paris Cit\'{e}\\
Universit\'{e}  Paris  Descartes, PARCC\\
European Hospital George Pompidou (HEGP-APHP)\\
LRI, INSERM U970-PARCC,   France\\
{\em ca@cuenod.net}\\

\noindent
Marianna Pensky\\
Department of Mathematics \\
University of Central Florida \\
Orlando FL 32816-1353, USA \\
{\em marianna.pensky@ucf.edu}\\

\noindent
Yves Rozenholc\\
Sorbonne Paris Cit\'{e}\\
Universit\'{e}  Paris  Descartes,\\
 MAP5, UMR CNRS 8145, France\\
{\em yves.rozenholc@parisdescartes.fr}

%%%%%%%%%%%%%%%%%%%%%%%%%%%%%%%%%%%%%%%%%%%%%%%%%%%%%%%%%%%%%%%%%%%%%%%%%%%%%%%%%%%%%%%%%%%%%%%%%%%%%%%%%%%%

\newpage

%%^^^^^^^^^^^^^^^^^^^^^^^^^^^^^^^^^^^^^^^^^^^^^^^^^^^^^^^^^^^^^^^^^^^^^^^^^^^^^^^^^^^^^^^^^^^^^^^^^^^^^^^^^^^^^^^^^^
%%%%%%%%%%%%%%%%%%%%%%%%%%%%%%%%%%%%%%%%%%%%%%%%%%%%%%%%%%%%%%%%%%%%%%%%%%%%%%%%%%%%%%%%%%%%%%%%%%%%%%%%%%%%%%%%%%%%
\appendix
\numberwithin{equation}{section}

\section{Supplementary materials}
\label{sec:appendix}

%%%%%%%%%%%%%%%%%%%%%%%%%%%%%%%%%%%%%%%%%%%%%%%%%%%%%%%%%%%%%%%%%%%%%%%%%%%%%%%%%%%%%%%%

\subsection{Introduction to theory of banded Toeplitz matrices}
\label{sec:ToeplitzIntro}

The proof of asymptotic optimality of the estimator $\hat{f}_{\widehat{m}}$
relies heavily on the theory of banded Toeplitz matrices developed in
B\"{o}ttcher  and Grudsky  (2000, 2005). In this subsection, we review some of the
facts about Toeplitz matrices which we shall use later.

Consider a sequence of numbers $\{ b_k \}_{k=-\infty}^\infty$  such that $\sum_{k=-\infty}^\infty |b_k| < \infty$.
An infinite Toeplitz matrix $T=T(b)$ is the matrix with elements $T_{i,j} = b_{i-j}$, $i,j=0,1, \ldots $.

Let ${\cal C} = \{z \in C: |z|=1 \}$ be the complex unit circle.
With each Toeplitz matrix $T(b)$ we can associate its symbol
\be \label{assocfun}
b(z) = \sum_{k=-\infty}^\infty b_k z^k, \ \ z \in {\cal C}.
\ee
Since, $\displaystyle{B(\theta)= b(e^{i\theta}) = \sum_{k=-\infty}^\infty b_k e^{i k \theta}}$,
numbers $b_k$ are Fourier coefficients of function $B(\theta)= b(e^{i\theta})$.

There is a very strong link between properties of a Toeplitz matrix $T(b)$
and function $b(z)$. In particular, if  $b(z) \neq 0$ for $z \in {\cal C}$
and $\wind (b) = J_b$, then $b(z)$ allows Wiener-Hopf factorization
$b(z) = b_{-} (z)\,  b_{+} (z)\,  z^{J_b}$  where   $b_+$ and $b_-$ have the following forms
$$
b_{-} (z)  =  \sum_{k=0}^\infty b^{-}_{-k} z^{-k}, \ \
b_{+} (z)  =  \sum_{k=0}^\infty b^{+}_{k} z^{k}
$$
(see Theorem 1.8 of B\"{o}ttcher  and Grudsky  (2005)).

If $T(b)$ is a lower triangular Toeplitz matrix, then
$b(z) \equiv b_{+} (z)$ with $b^+_k=b_k$.
%$b(z) \equiv b_{+} (z) = \displaystyle{\sum_{k=0}^\infty b_k z^k}$.
In this case,  the product of two Toeplitz matrices can be obtained by simply multiplying their symbols and
the inverse of a Toeplitz matrix can be obtained  by taking the  reciprocal
of function  $b_{+} (z)$:
\be \label{identity}
T(b_{+} d_{+}) = T(b_{+}) T(d_{+}),\ \ \
T^{-1}(b_{+}) = T(1/b_{+}).
\ee

Let $T_m (b) = T_m (b_{+}) \in R^{m \times m}$ be a banded lower triangular
Toeplitz matrix corresponding to the Laurent polynomial
$\displaystyle{b  (z) = \sum_{k=0}^{m-1} b_k z^k}$.

In practice, one usually use only finite, banded, Toeplitz matrices with
elements $T_{i,j}$, $i,j=0,1, \ldots, m-1$.  In this case,  only a finite  number
of coefficients $b_k$ do not vanish and function $b(z)$ in \fr{assocfun}
reduces to a  Laurent polynomial $\displaystyle{b(z) = \sum_{k=-J}^{K} b_k z^k}$,
 $z \in {\cal C}$,  where $J$ and $K$ are nonnegative integers, $b_{-J} \neq 0$ and  $b_{K} \neq 0$.
If $b(z) \neq 0$ for $z \in {\cal C}$, then $b(z)$ can be represented in a form
\be \label{Laurent_pol}
b(z) = z^{-J} b_{K} \prod_{j=1}^{J_0} (z-\mu_j) \prod_{k=1}^{K_0} (z - \nu_k) \ \ \mbox{with}\ \
|\mu_j|<1,\, |\nu_k|>1.
\ee
In this case, the winding number of $b(z)$ is $\wind (b) = J_0 - J$.

Let $T_m (b) = T_m (b_{+}) \in R^{m \times m}$ be a banded lower triangular
Toeplitz matrix corresponding to the Laurent polynomial
$\displaystyle{b  (z) = \sum_{k=0}^{m-1} b_k z^k}$.
If $b$ has no zeros on the complex unit circle ${\cal C}$   and $\wind(b) =0$, then, due to Theorem 3.7
 of  B\"{o}ttcher  and Grudsky  (2005),  $T(b)$ is invertible and
$\displaystyle{\lim_{m \rightarrow \infty} \sup \rho(T_m^{-1} (b)) < \infty}$.
Moreover, by Corollary 3.8,
\be \label{norm_converg}
\lim_{m \rightarrow \infty} \rho(T_m^{-1} (b)) = \rho(T^{-1} (b))
\ee

%%%%%%%%%%%%%%%%%%%%%%%%%%%%%%%%%%%%%%%%%%%%%%%%%%%%%%%%%%%%%%%%%%%%%%%%%%%%%%%%%%%%%%%%%%%%%%%%%%%%%%%55

\subsection{Relation between $\rho_m^2$ and $v_m^2$ }
\label{sec:relation}

In order to apply the theory surveyed above, we first need to
examine function $b(z)$ associated with the infinite lower triangular Toeplitz matrix
$\bG$  defined in Lemma \ref{lem:triangular_system}
the Laurent polynomial associated with its banded version
$\bG_m$. It turns out that $b(z)$ can be expressed via the Laplace transform $G(s)$
of the kernel $g(t)$. In particular, the following statement holds.

\begin{lemma} \label{lem:Toeplitz1}
Consider a sequence $\{ b_k \}_{k=0}^\infty$ with elements $b_0=g^{(0)}$
and $b_k = g^{(k)}  - g^{(k-1)}$, $k=1,2, \ldots$ where $g^{(k)}$ are Laguerre  coefficients of
the kernel $g$ in \fr{eq:model}. Then,   $b_k$, $k \geq 0$,  are Fourier coefficients of the function
\be \label{eq:Fourier_tr}
b(e^{i \theta}) =   G \lkr \frac{a(1+e^{i \theta})}{(1-e^{i \theta})} \rkr = \sum_{k=0}^\infty b_k e^{i \theta k},
\ee
where $G(s)$ is the Laplace transform of the kernel $g(x)$.
\end{lemma}

\noindent{\bf Proof. }
% \noindent{\bf Proof  of Lemma  \ref{lem:Toeplitz1}. }
To prove this statement, we shall follow the theory of Wiener-Hopf integral equations
described in Gohberg and Feldman (1974).
Denote Fourier transform of a function $p(x)$ by
$ \hat{p} (\omega) = \displaystyle{\int_{-\infty}^\infty e^{i \omega x} p(x) dx}$
and observe that
$$
\hat{\phi}_k(\om) = (-1)^k \sqrt{2a} \frac{(a + i \om)^k}{(a- i \om)^{k+1}}.
$$
Therefore, elements of the infinite Toeplitz matrix $\bG$ in Lemma \ref{lem:triangular_system}
are generated by the sequence $b_j$, $j \geq 0$, where
\beqn
b_j & = & (2a)^{-1/2} (g^{(j)} - g^{(j-1)}) =
\frac{1}{2\pi}\ \int_{-\infty}^\infty \hat{g} (\om)
\overline{[\hat{\phi}_j(\om) - \hat{\phi}_{j-1} (\om)]} d\om \nonumber \\
& = &  \frac{a}{\pi} \int_{-\infty}^\infty \hat{g}(\om)
\lkr \frac{i \om -a}{i \om +a} \rkr^j \frac{d\om}{a^2 + \om^2},\ \ j=0,1, \ldots.
\label{Toep_seq1}
\eeqn
 Note that $|(i \om -a)/(i \om +a)|=1$, so that we can use the following substitution in the integral \fr{Toep_seq1}:
$$
\frac{i \om -a}{i \om +a}  = e^{-i \theta} \  \Longrightarrow \ \om = \frac{a(e^{i\theta} + 1)}{i(e^{i\theta} - 1)}
= \frac{a \sin \theta}{\cos \theta -1}, \ \ 0 \leq \theta \leq 2\pi.
$$
Simple calculations show that
$$
b_j  = \frac{1}{2\pi}\ \int_{0}^{2\pi} \hat{g} \lkr  \frac{a(e^{i\theta} + 1)}{i (+e^{i\theta} - 1)} \rkr
e^{-i\theta j} d\theta,
$$
so that $b_j$, $j \in \ZZ$, are Fourier coefficients of the function
$$
B(\theta) = b(e^{i \theta})=  \hat{g} \lkr  \frac{a(e^{i\theta} + 1)}{i( e^{i\theta} - 1)} \rkr.
$$
Now, let us show that $b_j=0$ for $j<0$. Indeed, if $j=-k$, $k > 0$, then
\beqns
b_j & = &   \frac{a}{\pi} \int_{-\infty}^\infty \hat{g}(\om)
\lkr \frac{i \om + a}{i \om -a} \rkr^k \frac{d\om}{a^2 + \om^2}
= \frac{a}{\pi} \int_{-\infty}^\infty \hat{g}(\om)
\lkr \frac{i (-\om) - a}{i (-\om) + a} \rkr^k \frac{d\om}{a^2 + \om^2} \\
& = &
\frac{1}{2\pi}\ \int_{-\infty}^\infty \hat{g} (\om)
\overline{[\hat{\phi}_j(-\om) - \hat{\phi}_{j-1} (-\om)]} d\om
=  \int_{-\infty}^\infty g(x) \lkv \phi_j(-x) - \phi_{j-1}(-x) \rkv  dx = 0
\eeqns
since $g(x)=0$ if $x<0$ and $\phi_k(-x) = 0$ if $x>0$.
Hence, function $B(\theta) = b(e^{i \theta})$ has only coefficients
$b_j$, $j \geq 0$, in its Fourier series.
Now, to complete the proof, one just needs to note that $G(s) = \hat{g}(i s)$
for any $s$ such that Laplace transform $G(s)$ of $g$ exists.
\\

For any function $w(z)$ with an argument on a unit circle ${\cal C}$ denote
$$
\|w\|_{circ} = \displaystyle{\max_{|z|=1} w(z)}.
$$
The following lemma \ref{lem:rho_v} shows that indeed  $\rho_m^2  \log m  = o(v_m^2)$   as
$m \rightarrow \infty$.
\\

\begin{lemma} \label{lem:rho_v}
Let $b(z)$ be given by \fr{eq:Fourier_tr}, i.e., $b(z) = G (a(1+z)/(1-z))$, $|z|=1.$
Denote
\be \label{eq:q}
w(z) = (1-z)^{-r} b(z),\ \ w^{-1} (z) = (1-z)^{r} b^{-1}(z),\ \
|z|=1.
\ee
Then, under assumptions {\bf (A1)}--{\bf (A3)}, $w(z)$ and $w^{-1} (z)$ have  no zero on the complex
unit circle and,  for $m$   large enough,  one has
\begin{eqnarray}
\frac{\lambda_1}{2(r!)^2} \, \lkr \|w\|_{circ} \rkr^{-2}\  m^{2r} &\leq &  \rho_m^2    \leq  \  v_m^2  \leq 2  \lambda_2\, \|w^{-1} \|_{circ}^2\   m^{2r},
\label{eq:vm_normratio}
 \end{eqnarray}
% \begin{eqnarray}
% \frac{\lambda_1}{2^{4r} [(r-1)!] } \lkr \|w\|_{circ} \rkr^{-2}\  m^{2r} &\leq& v_m^2 \leq 2  \lambda_2\, \|w^{-1} \|_{circ}^2\   m^{2r},
% \label{eq:vm_norm}\\
% \frac{\lambda_1}{2(r!)^2} \, \lkr \|w\|_{circ} \rkr^{-2}\  m^{2r} &\leq &  \rho_m^2    \leq  \  v_m^2,  \label{eq:rhom_vm_ratio}
% \end{eqnarray}
where $\rho_m^2$ and $v_m^2$ are defined in \fr{eq:vm_rhom}, $\lambda_1$ and $\lambda_2$ are given by \fr{eigenvalues}. 
%and
%$C(r,w)$ is an absolute constant which depends only on $w$ and $r$:
%$$
%C(r,w) = 2^{4r+1}\, [(r-1)!]^2\, \lkr \|w\|_{circ}\, \|w^{-1} \|_{circ} \rkr^2\ \lambda_2/ \lambda_1.
%$$
\end{lemma}

\noindent{\bf Proof  of Lemma  \ref{lem:rho_v}. }
Let us first find upper and  lower bounds on $\| \bG_m^{-1} \|_2^2 = \Tr(\bG_m^{-T} \bG_m^{-1})$ and
  $\| \bG_m^{-1} \|^2 = \lambda_{\max} (\bG_m^{-T} \bG_m^{-1})$.
For this purpose,  examine the function
$$
b(z)  = \hat{g} \lkr  \frac{a(z + 1)}{i (z - 1)} \rkr = G \lkr  \frac{a(z + 1)}{ 1-z } \rkr,\ \ |z|=1.
$$
Denote $y = a(z + 1)/(1-z)$, so that $z = (y-a)/(y+a)$ and  $G(y) = b((y-a)/(y+a))$.

Let us show that, under Assumptions {\bf (A1)}-{\bf (A3)}, $b(z)$ has a zero  of order $r$ at $z=1$ and
all other zeros of $b(z)$ lie outside the unit circle.

For this purpose, assume that $y=\alpha + i \beta$ is a zero of $G$, i.e. $G(\alpha + i \beta) =0$.
Simple calculus yields
$$
\left| \frac{y-a}{y+a} \right|^2 = 1 - \frac{4 \alpha a}{(\alpha +a)^2 + \beta^2},
$$
so that $|z| =  |(y-a)/(y+a)| \leq 1$ iff $\alpha \geq 0$. But, by Assumption {\bf (A2)}, $G(y)$ has no zeros with
nonnegative real parts, so that $\alpha <0$ and $|z| =  |(y-a)/(y+a)| > 1$. Therefore, all zeros of $b(z)$,
which correspond to finite  zeros of $G$, lie outside the complex unit circle ${\cal C}$.

 Assumptions  {\bf (A1)}, {\bf (A2)} and properties of Laplace transform imply that
$G(s) = s^{-r} (B_r + G_r(s))$ where $G_r(s)$ is the Laplace transform of
$g^{(r)} (t)$. Hence,
$$
 \lim_{Re\   s \rightarrow \infty} s^j G(s) = \lfi
\begin{array}{ll}
0, & \mbox{if}\ \ j=0, ..., r-1,\\
B_r \ne 0, &  \mbox{if}\ \ j=r,
%\\ \ne 0
\end{array} \right.
$$
so that $y = \infty + i \beta$ is zero of order $r$ of $G(y)$.
Since $\displaystyle{\lim_{Re\  y \rightarrow \infty}  (y-a)/(y+a) =1}$,
$b(z)$ has zero of order $r$ at $z=1$.

Then, $b(z)$ can be written as $b(z) = (1-z)^r w(z)$ where $w(z)$ is defined by formula \fr{eq:q}
and all zeros of $w(z)$ lie outside the complex unit circle. Therefore, $w(z)$ can be written as
\be \label{w_represent}
w(z) = C_w \prod_{j=1}^N (z - \zeta_j),\ \ 0 \leq N \leq \infty,\ |\zeta_j|>1,
\ee
where $C_w$ is an absolute constant. Since $b(z)$ does not contain any negative powers of $z$
in its representation, $J_0 =0$ and $J=0$ in \fr{Laurent_pol}   and, consequently, $\wind(w) =0$.
Also, by \fr{identity} and \fr{eq:q}, one has
$T^{-1}(b) = T(b^{-1})$ where $b^{-1}(z) =   w^{-1} (z) (1-z)^{-r}$.

Now, recall that $ \| \bG_m^{-1} \|_2^2 =   \|T_m(b^{-1})\|_2^2$ and
$\rho^2(\bG_m^{-1})  =  \rho^2(T_m(b^{-1}))$.
 Using relation between Frobenius and spectral norms $\|\bA_1 \bA_2 \|_2 \leq \|\bA_1\|_2 \rho(\bA_2)$
for any matrices $\bA_1$ and $\bA_2$
(see, e.g., B\"{o}ttcher  and Grudsky  (2000), page 116), obtain
\begin{equation} \label{frob}
\|T_m(b^{-1})\|_2 \leq \|T_m((1-z)^{-r})\|_2  \rho(T_m(w^{-1})), \;\;
 \rho(T_m(b^{-1}))  \leq \rho(T_m((1-z)^{-r}))   \rho(T_m(w^{-1})),
\end{equation}
\begin{equation}
 \|T_m((1-z)^{-r})\|_2 \leq \|T_m(b^{-1})\|_2  \rho(T_m(w)), \;\;
\rho(T_m((1-z)^{-r}))  \leq \rho(T_m(b^{-1}))   \rho(T_m(w)).  \label{eigen}
\end{equation}

Note that (see B\"{o}ttcher  and Grudsky  (2005), page 13)
$$
\lim_{m \rightarrow \infty}  \rho(T_m(w^{-1})) =  \|w^{-1} \|_{circ},\ \
\lim_{m \rightarrow \infty}  \rho(T_m(w )) = \|w\|_{circ},
$$
Also, due to representation \fr{w_represent}, both    $w$ and $w^{-1}$ are bounded,
and, therefore, $0 < \|w^{-1} \|_{circ} < \infty$ and   $0 < \|w \|_{circ} < \infty$. Denote
\be \label{def:nu_f_nu_s}
\nu_f (m) = \|T_m((1-z)^{-r})\|_2,\ \ \nu_s (m) = \rho(T_m((1-z)^{-r})).
\ee
Then, it follows from \fr{norm_converg}, \fr{frob} and \fr{eigen}   that, for $m$ large enough,
\beqn
0.5 \lkr \|w\|_{circ} \rkr^{-2}\  \nu_f^2 (m) & \leq & \|T_m(b^{-1}) \|_2^2 \leq 2 \|w^{-1} \|_{circ}^2\  \nu^2_f(m),
\label{Frobenius1}\\
0.5 \lkr \|w\|_{circ} \rkr^{-2}\  \nu_s^2 (m) & \leq & \rho^2(T_m(b^{-1}))  \leq 2 \|w^{-1} \|_{circ}^2\  \nu^2_s(m).
\label{spectral1}
\eeqn

In order to finish the proof, we need to evaluate $\nu^2_f(m)$ and  $\nu^2_s(m)$ and also to
derive a relation between $v_m^2$, $\rho_m^2$, $\|T_m(b^{-1}) \|_2^2$ and $\rho^2(T_m(b^{-1}) )$.
The first task is accomplished by the following lemma.

\begin{lemma} \label{lem:nu_f_nu_s}
Let $\nu_f(m)$ and $\nu_s (m)$ be defined in \fr{def:nu_f_nu_s}.
Then,
\beqn
\nu^2_f(m) &\leq & m^{2r}, \label{Frobenius2}\\
(r!)^{-2} m^{2r} & \leq & \nu^2_s(m). \label{spectral2}
\eeqn
% \beqn
%  2^{-(4r-1)} [(r-1)!]^{-2} m^{2r} & \leq & \nu^2_f(m) \leq m^{2r}, \label{Frobenius2}\\
% (r!)^{-2} m^{2r} & \leq & \nu^2_s(m) \leq m^{2r}. \label{spectral2}
% \eeqn
\end{lemma}

Proof of Lemma \ref{lem:nu_f_nu_s}  is given in Section \ref{sec:supplementary}.
\\

Now, to complete the proof, note that due to relation between Frobenius and spectral norms
\beqns
%\|T_m(b^{-1}) \|_2^2 & = &  \Tr (\bG_m^{-T} \bG_m^{-1})  =
%\Tr (\bOm_m^{-1} \bQ_m) \leq   \lambda_1^{-1} v_m^2,\\
v_m^2  & = & \Tr (\bOm_m \bG_m^{-T} \bG_m^{-1}) \leq \lambda_2 \|\bG_m^{-1} \|_2^2
= \lambda_2  \|T_m(b^{-1}) \|_2^2,  \\
\rho^2(T_m(b^{-1}) ) & = &  \lambda_{{\rm max}}(\bG_m^{-T} \bG_m^{-1})  
 \leq   \lambda_1^{-1} \rho_m^2,
%\rho_m^2 & \leq & \lambda_{\max} (\bOm_m) \lambda_{\max}(\bG_m^{-T} \bG_m^{-1}) \leq \lambda_2 \rho^2(\bG_m^{-1} ) = \lambda_2  \rho^2(T_m(b^{-1})),
\eeqns
so that
\be \label{Frobenius_spectral}
\rho_m^2  \geq \lambda_1  \rho^2(T_m(b^{-1})), \,
 v_m^2 \leq \lambda_2 \|T_m(b^{-1}) \|_2^2.
\ee
% \be \label{Frobenius_spectral}
% \rho_m^2  \geq \lambda_1  \rho^2(T_m(b^{-1})), \ \
% \lambda_1 \|T_m(b^{-1}) \|_2^2 \leq v_m^2 \leq \lambda_2 \|T_m(b^{-1}) \|_2^2.
% \ee
Combination of \fr{Frobenius1} -- \fr{Frobenius_spectral} and
Lemma \ref{lem:nu_f_nu_s} complete the proof.

%%%%%%%%%%%%%%%%%%%%%%%%%%%%%%%%%%%%%%%%%%%%%%%%%%%%%%%%%%%%%%%%%%%%%%%%%%%%%%%%%%%%%%%%%%%%%%%%%%%%%%%%%%%%%%5

%%%%%%%%%%%%%%%%%%%%%%%%%%%%%%%%%%%%%%%%%%%%%%%%%%%%%%%%%%%%%%%%%%%%%%%%%%%%%%%%%%%%%%%%%%%%%%%%%%%%%%%%%%%%%%%%%

\subsection {Proofs of supplementary Lemmas }
\label{sec:supplementary}

\noindent{\bf Proof of Lemma  \ref{lem:large_deviation}. }

The proof of Lemma \ref{lem:large_deviation} has two steps. The first one is the application of a $\chi^2$-type deviation inequality. The second step consists of integrating this deviation inequality.

 The $\chi^2$-inequality is formulated as follows. 
In the Gaussian case, it is stated in Laurent and Massart  (2000), and  improved by Gendre (see Lemma 8.2 of Gendre  (2009)).
In the sub-Gaussian case, it is given in Rudelson and Vershynin~(2013), Theorem 2.1.
 Let $\bA$   be a $p\times p$ matrix $\bA\in {\mathbb M}_p({\mathbb R})$ and $\bzeta$ be a vector of sub-Gaussian random variables. Then, for any $x>0$,
\begin{equation}\label{gendre1}
{\mathbb P}\lkr \|\bA\bzeta\|^2\geq \|\bA\|_2^2 + 2\sqrt{\|\bA\|_2^2 \rho^2(\bA) x} + \rho^2(\bA) x \rkr \leq 2e^{-x/\kappa}.
\end{equation}
In the Gaussian case, namely,  for $\bzeta$ a standard Gaussian vector, we have $\kappa=1$.

%and \begin{equation}\label{gendre2} {\mathbb P} (\|\bA\bzeta\|^2\leq v_A^2 - 2\sqrt{\rho^2(\bA)v_A^2 x})\leq e^{-x}.\end{equation}

Now,  recall that for $\bt\in S_m+S_{m'}=S_{m^*}$ where   $m^*=m\vee m'$, one has
$$
\langle \bt, \bG_M^{-1} (\Phi_M^T\Phi_M)^{-1}\Phi_M^T\vec\varepsilon \; \rangle =
\sqrt{\frac{T}n} \langle [{\mathbf t}]_{m^*}, {\mathbf A}_{m^*} \vec \varepsilon \; \rangle
$$
where we recall that $[{\mathbf t}]_{m*}$ is the $m^*$-dimensional vector formed by  the first $m^*$ coordinates of $\bt$ and ${\mathbf A}_m$ is defined by (\ref{def:Am}). Moreover,
$$
\sup_{\bt\in S_m+S_{m'}, \|\bt\|=1}  \langle \bt, \bG_{M}^{-1}(\Phi_M^T\Phi_M)^{-1}\Phi_M^T\vec\varepsilon \;\rangle^2
= \frac{ T}n \|{\mathbf A}_{m^*} \vec \varepsilon \; \|^2.
$$
Thus, it follows from (\ref{gendre1}) that
\begin{equation}\label{gendre3}
  {\mathbb P} \left( \|{\mathbf A}_{m^*} \vec \varepsilon \; \|^2\geq v_{m^*}^2 +
2\sqrt{\rho_{m^*}^2v_{m^*}^2x} +\rho_{m^*}^2 x \right)\leq 2e^{-x/\kappa}.
\end{equation}
One has $2\sqrt{\rho_{m^*}^2v_{m^*}^2x}\leq v_{m^*}^2 +  \rho_{m^*}^2 x$ so that
$$
{\mathbb P}\left(\|{\mathbf A}_{m^*} \vec \varepsilon \; \|^2 \geq 2 v_{m^*}^2 + 2 \rho_{m^*}^2 x\right)\leq 2 e^{-x/\kappa}.
$$
Therefore, using definition \fr{eq:tau} of $\tau(m,m')$, obtain
\begin{eqnarray*}
&&{\mathbb E}\left(\sup_{\bt\in S_m+S_{m'}, \|\bt\|=1}  \langle \bt, \bG_M^{-1}
(\Phi_M^T\Phi_M)^{-1}\Phi_M^T\vec\varepsilon \; \rangle ^2 -\tau(m,m')\right)_+
= {\mathbb E}\left( \frac{T}n  \|{\mathbf A}_{m^*} \vec \varepsilon \; \|^2-\tau(m,m')\right)_+ \\
&\leq & \frac{T}n \int_0^{+\infty} {\mathbb P}\left(\|{\mathbf A}_{m^*} \vec \varepsilon \; \|^2-
\lkv 2 v_{m^*}^2 + 2 \log[(m^*\rho_{m^*}/\rho_1)^{2\kappa}]\rho_{m^*}^2 \rkv \geq  \xi \right) d\xi .
\end{eqnarray*}
Changing variables
$$
2  \log[(m^*\rho_{m^*}/\rho_1)^{2\kappa}]\rho_{m^*}^2 +\xi= 2 \rho_{m^*}^2 x
 $$
and application of  (\ref{gendre3}) yield

\begin{eqnarray*}
{\mathbb E}\left(\sup_{\bt\in S_m+S_{m'}, \|\bt\|=1} \langle \bt, \bG_{M}^{-1} \Phi_M^T\Phi_M)^{-1}\Phi_M^T\vec\varepsilon \; \rangle ^2 -\tau(m,m')\right)_+
&\leq & 4  \frac Tn\rho_{m^*}^2\int_{\log[(m^*\rho_{m^*}/\rho_1)^{2\kappa}]}^{+\infty} e^{-x/\kappa}dx \\
&=& 4\frac Tn \rho_1^2 (m^*)^{-2}.
\end{eqnarray*}

Thus we obtain
\begin{eqnarray*}
&& \EE \lkv \sup_{\bt \in S_{m \vee \whm}}  \langle \bt, \bG_M^{-1} \Phi_M^T\Phi_M)^{-1}\Phi_M^T\vec\varepsilon \; \rangle^2 - \tau(m,\whm) \rkv_+ \\ & \leq &
\sum_{m' \in  \mathM_n} \EE \left(\sup_{\bt\in S_m+S_{m'}, \|\bt\|=1}
\langle \bt, \bG_M^{-1} \Phi_M^T\Phi_M)^{-1}\Phi_M^T\vec\varepsilon \; \rangle^2 - \tau(m,m') \right)_+
\end{eqnarray*}
and
\begin{eqnarray*}
&& \sum_{m'\in {\mathcal M}_n} {\mathbb E}\left(\sup_{\bt\in S_m+S_{m'}, \|\bt\|=1}
\langle \bt, \bG_{M}^{-1} \Phi_M^T\Phi_M)^{-1}\Phi_M^T\vec\varepsilon \; \rangle ^2 - \tau(m,m')\right)_+
\\ &\leq &
4 \rho_1^2 \frac{T}n \sum_{m'\in {\mathcal M}_n} (m\vee m')^{-2} \\
&\leq &  4 \rho_1^2\frac{T}n \left( \sum_{m'=1}^m m^{-2}  + \sum_{m'> m} (m')^{-2}\right)\\
&\leq &  4\rho_1^2 \frac{T}n \left( m^{-1}  + \int_m^{+\infty} \frac{dx}{x^2} \right)
=8\rho_1^2 \frac{T}{n m},
\end{eqnarray*}
which concludes the proof. $\Box$\\

%%%%%%%%%%%%%%%%%%%%%%%%%%%%%%%%%%%%%%%%%%%%%%%%%%%%%%%%%%%%%

\noindent{\bf Proof of Lemma  \ref{lem:nu_f_nu_s}. }
Note that, by formula 1.110 of Gradshtein and  Ryzhik  (1980),
$$
(1-z)^{-r} = \sum_{j=0}^\infty {r+j-1 \choose j} z^j,
$$
so that, by definition of Frobenius norm,
\beqns
% \nu_f^2 (m) =
\|T_m((1-z)^{-r})\|^2_2 & =&  m + (m-1) {r \choose 1}^2 + (m-2) {r+1  \choose 2}^2 +
\ldots + {r+m-2 \choose m-1}^2 \\
& = & \sum_{j=0}^{m-1} {r+j-1 \choose j}^2 (m-j),\\
\rho ( T_m((1-z)^{-r}))  & =& \max_{|z|=1} \left| \sum_{j=0}^{m-1} {r+j-1 \choose j} z^j \right|
= \sum_{j=0}^{m-1} {r+j-1 \choose r-1}.
\eeqns
If $r=1$, then
$$
\sum_{j=0}^{m-1} {r+j-1 \choose j}^2 (m-j) = \sum_{j=0}^{m-1}   (m-j) = \frac{m(m+1)}{2}.
$$
If $r \geq 2$, then
$$
\frac{j^{r-1}}{(r-1)!} \leq {r+j-1 \choose j} = \frac{(r-1+1) \ldots (r-1+j)}{(r-1)!}
\leq (j+1)^{r-1},
$$
so that, for $m \geq 4$,
\beqns
 \nu_f^2 (m) & = & \|T_m((1-z)^{-r})\|^2_2   \leq    m^{2r}, \\
% \nu_f^2 (m)  & \geq & \sum_{j=m/4}^{3m/4} \frac{j^{2r-2}}{[(r-1)!]^2} (m-j) \geq
%\frac{m^{2r} 2^{-(4r-1)}}{[(r-1)!]^2},
\eeqns
which proves validity of \fr{Frobenius2}. To show that \fr{spectral2} holds,
observe that, by formula 0.151.1 of Gradshtein and  Ryzhik  (1980),
$$
\sum_{j=0}^{m-1} {r+j-1 \choose r-1} = {r+m-1 \choose r},\ \    \frac{m^r}{r!} \leq {r+m-1 \choose r} \leq m^r.
$$

%%%%%%%%%%%%%%%%%%%%%%%%%%%%%%%%%%%%%%%%%%%%%%%%%%%%%%%%%%%%%%%%%%%%%%%%%%%%%%%%%%%%%%%%

%%%%%%%%%%%%%%%%%%%%%%%%%%%%%%%%%%%%%%%%%%%%%%%%%%%%%%%%%%%%%%%%%%%%%%%%%%%%%%%%%%%%%%%%%%%%%%%%%%%%%%%%%%%%%%%%%

 \section{Simulation tables}

 Table \ref{tab:risks}   provides the averages and their standard deviations (in italic) of the $ISE(\widehat f_{\mathrm{LAG}})$ 
computed over 400 simulation runs.\\
\begin{table}[!h] \centering\small
\[
\begin{array} {lccc|cc|cc|cc|cc}
 & & \multicolumn{2}{c}{g_1} & \multicolumn{2}{c}{g_2} & \multicolumn{2}{c}{g_3} & \multicolumn{2}{c}{g_4} & \multicolumn{2}{c}{g_5} \\\cline{3-12}
 & n & 100 & 250 & 100 & 250 & 100 & 250 & 100 & 250 & 100 & 250\\\cline{2-12}
\multirow{10}{*}{$f_1$} & \multirow{2}{*}{$i=1$} & 173 & 39.9 & 5.50 & 2.33 & 31.8 & 11.9 & 182 & 123 & 201 & 175 \\ 
 &  & \it 50.4 & \it 52.4 & \it 7.07 & \it 2.50 & \it 46.9 & \it 13.0 & \it 50.9 & \it 67.3 & \it 80.4 & \it 52.4 \\ 
 & \multirow{2}{*}{$i=2$} & 21.7 & 14.9 & 1.35 & 0.69 & 7.94 & 3.35 & 35.4 & 12.7 & 110 & 20.5 \\ 
 &  & \it 30.5 & \it 9.02 & \it 2.07 & \it 0.87 & \it 11.7 & \it 6.30 & \it 54.1 & \it 14.3 & \it 80.0 & \it 24.3 \\ 
 & \multirow{2}{*}{$i=3$} & 6.18 & 10.1 & 0.65 & 0.35 & 2.18 & 0.87 & 8.42 & 9.89 & 13.6 & 10.6 \\ 
 &  & \it 9.22 & \it 6.91 & \it 0.87 & \it 0.29 & \it 4.91 & \it 1.54 & \it 12.4 & \it 9.86 & \it 20.1 & \it 6.88 \\ 
 & \multirow{2}{*}{$i=4$} & 10.9 & 5.15 & 0.14 & 0.089 & 0.60 & 0.28 & 1.99 & 6.20 & 4.93 & 22.2 \\ 
 &  & \it 5.63 & \it 4.23 & \it 0.13 & \it 0.080 & \it 1.01 & \it 0.64 & \it 3.36 & \it 4.29 & \it 14.0 & \it 27.6 \\ 
 & \multirow{2}{*}{$i=5$} & 2.05 & 1.67 & 0.050 & 0.048 & 0.24 & 0.095 & 3.27 & 2.16 & 10.8 & 4.79 \\ 
 &  & \it 2.01 & \it 0.50 & \it 0.033 & \it 0.043 & \it 0.55 & \it 0.18 & \it 9.72 & \it 2.77 & \it 10.7 & \it 3.39 \\ \hline
\multirow{10}{*}{$f_2$} & \multirow{2}{*}{$i=1$} & 252 & 62.7 & 20.7 & 20.5 & 7.18 & 3.34 & 896 & 215 & 144 & 389 \\ 
 &  & \it 65.2 & \it 42.0 & \it 8.79 & \it 7.07 & \it 14.0 & \it 4.59 & \it 637 & \it 148 & \it 290 & \it 115 \\ 
 & \multirow{2}{*}{$i=2$} & 194 & 20.1 & 10.5 & 13.3 & 1.77 & 0.77 & 871 & 321 & 285 & 29.0 \\ 
 &  & \it 239 & \it 14.3 & \it 6.07 & \it 3.83 & \it 5.11 & \it 1.19 & \it 685 & \it 174 & \it 202 & \it 12.5 \\ 
 & \multirow{2}{*}{$i=3$} & 160 & 17.3 & 5.75 & 8.3 & 0.46 & 0.20 & 149 & 252 & 179 & 34.5 \\ 
 &  & \it 37 & \it 4.39 & \it 3.22 & \it 3.69 & \it 1.37 & \it 0.30 & \it 176 & \it 72.8 & \it 53.6 & \it 98.8 \\ 
 & \multirow{2}{*}{$i=4$} & 227 & 73.6 & 3.47 & 3.5 & 0.14 & 0.056 & 111 & 12.5 & 242 & 59.1 \\ 
 &  & \it 26.6 & \it 74.4 & \it 2.91 & \it 1.7 & \it 0.54 & \it 0.065 & \it 39.2 & \it 64.5 & \it 127 & \it 73.3 \\ 
 & \multirow{2}{*}{$i=5$} & 257 & 44.7 & 2.35 & 1.59 & 0.23 & 0.019 & 218 & 62 & 224 & 51.4 \\ 
 &  & \it 41.4 & \it 31.5 & \it 3.93 & \it 0.39 & \it 0.65 & \it 0.023 & \it 82.1 & \it 74.3 & \it 61.8 & \it 19.5 \\ \hline
\multirow{10}{*}{$f_3$} & \multirow{2}{*}{$i=1$} & 46.1 & 22.2 & 5.33 & 2.59 & 27.9 & 23.0 & 72.2 & 31.7 & 165 & 45.2 \\ 
 &  & \it 51.3 & \it 13.3 & \it 5.19 & \it 3.03 & \it 26.3 & \it 23.3 & \it 106 & \it 25.9 & \it 238 & \it 43.1 \\ 
 & \multirow{2}{*}{$i=2$} & 21.2 & 13.0 & 1.46 & 0.77 & 16.2 & 5.09 & 27.3 & 16.7 & 45.2 & 20.4 \\ 
 &  & \it 15.8 & \it 3.15 & \it 1.55 & \it 0.82 & \it 16.5 & \it 6.04 & \it 27.0 & \it 8.49 & \it 57.4 & \it 12.3 \\ 
 & \multirow{2}{*}{$i=3$} & 14.0 & 5.60 & 0.46 & 0.21 & 3.77 & 1.43 & 18.1 & 7.64 & 23.5 & 16.3 \\ 
 &  & \it 5.78 & \it 7.02 & \it 0.36 & \it 0.20 & \it 5.01 & \it 1.75 & \it 11.8 & \it 8.24 & \it 24.4 & \it 10.1 \\ 
 & \multirow{2}{*}{$i=4$} & 9.94 & 2.01 & 0.108 & 0.056 & 1.00 & 0.55 & 4.89 & 1.76 & 14.6 & 2.85 \\ 
 &  & \it 8.68 & \it 2.50 & \it 0.14 & \it 0.055 & \it 1.33 & \it 0.61 & \it 7.03 & \it 2.10 & \it 13.7 & \it 3.38 \\ 
 & \multirow{2}{*}{$i=5$} & 7.62 & 1.00 & 0.025 & 0.015 & 0.50 & 0.17 & 2.50 & 0.51 & 7.70 & 0.75 \\ 
 &  & \it 3.39 & \it 0.89 & \it 0.019 & \it 0.018 & \it 0.56 & \it 0.149 & \it 2.81 & \it 0.46 & \it 5.59 & \it 0.77 \\ 
\end{array}
\]
\caption{\small Empirical (over 400 simulation runs) Mean Integrated Square Errors ($\times 10^4$) and {\it standard deviations in italic}
for  kernels $g_j$ ($j=1,\ldots,5$), unknown functions $f_1$ to $f_3$,  $n=100$ and $n=250$, and, for the noise level equals to
$\sigma_0(g_j)/2^i$, $i=1,\ldots,5$.
% For each empirical risk, 400 simulations has been ran starting with a seed equal to 8122012 in order to allow reproducible computations.
\label{tab:risks}}
\end{table}

%%%%%%%%%%%%%%%%%%%%%%%%%%%%%%%%%%%%%%%%%%%%%%%%%%%%%%%%%%%%%%%%%%%%%%%%%%%%%%%%%%%%%%%%%%%%%%%%%%%%%%%%%%%%5
%%^^^^^^^^^^^^^^^^^^^^^^^^^^^^^^^^^^^^^^^^^^^^^^^^^^^^^^^^^^^^^^^^^^^^^^^^^^^^^^^^^^^^^^^^^^^^^^^^^^^^^^^^^^^^^^^^^^

%%%%%%%%%%%%%%%%%%%%%%%%%%%%%%%%%%%%%%%%%%%%%%%%%%%%%%%%%%%%%%%%%%%%%%%%%%%%%%%%%%%%%%%%%%%%%%%%%%%%%%%%%%%%

\end{document}